\RequirePackage{fix-cm}
\documentclass[natbib,smallextended]{svjour3}       % onecolumn (second format)
\bibpunct{[}{]}{;}{n}{}{,} % to get "[numbered]" references from natbib

\smartqed  % flush right qed marks, e.g. at end of proof
\usepackage{graphicx}
\usepackage{color}
\usepackage{amsmath,amssymb}
 \journalname{my journal}
\newcommand {\nc} {\newcommand}
\nc {\beq} {\begin{eqnarray}}
\nc {\eeq} {\nonumber \end{eqnarray}}
\nc {\eeqn} [1] {\label{#1} \end{eqnarray}}
\nc {\eol} {\nonumber \\}
\nc {\rref} [1] {(\ref{#1})}
\nc{\ETAL} {\mbox{\sl et al.}}
\nc {\vR} {\mbox{$\ve{R}$}}
\nc {\ve} [1] {\mbox{\boldmath $#1$}}
\nc {\la} {\mbox{$\langle$}}
\nc {\ra} {\mbox{$\rangle$}}
\nc {\cL} {\mbox{${\cal L}$}}
\nc {\dem} {\mbox{$\frac{1}{2}$}}
\nc {\arrow} [2] {\mbox{$\mathop{\rightarrow}\limits_{#1 \rightarrow #2}$}}
\nc {\wiggle} [2] {\mbox{$\mathop{\sim}\limits_{#1 \rightarrow #2}$}}
\nc {\red}[1] {\textcolor{red}{#1}}
\begin{document}

\title{Natural and Dyson orbitals  in small  helium drops.
}

\titlerunning{Natural and Dyson orbitals in small helium drops 
%: application to  sub-threshold halo states in helium drops
}        % if too long for running head

\author{N.K. Timofeyuk
}


\institute{ N.K. Timofeyuk \at
              School of Mathematics and Physics, University of Surrey, Guildford,
Surrey GU2 7XH,  UK \\
\email{ n.timofeyuk@surrey.ac.uk}
}

\date{Received: date / Accepted: date}

\maketitle

\begin{abstract}

The natural and Dyson orbitals are studied for small helium drops comprising 5 to 20 helium atoms interacting via a soft two-body gaussian potential. The wave functions of these drops have been obtained in the hyperspherical cluster model (HCM) which provides a correct description of the single-particle behaviour  at large separations from the  system. The natural orbitals are obtained from diagonalization of  the nonlocal one-body density matrix, while Dyson orbitals are constructed by direct overlap of the wave functions of two drops differing by one boson. This overlap converges with increasing  basis of the HCM.
The  shapes and occupancies of the natural orbitals as well as their link  to Dyson overlaps and evolution with increasing number of atoms are discussed. Both natural and Dyson orbitals can be used to represent the density of the system. However, the natural orbitals representation is demonstrated to be superior.  With increasing boson numbers the difference between Dyson and natural orbitals becomes less prominent and it is expected to disappear in infinitely large systems of identical bosons.

%Formal expressions for nonlocal density matrix elements were developed within the HCM to enable calculations of natural orbitals. The  shapes and occupancies of these orbitals are discussed together with their link to Dyson overlaps between the wave functions of two drops differing by one boson. The most striking feature revealed in this work  is  abnormality in achieving correct asymptotic decrease of Dyson orbitals corresponding to  population of excited states via one boson removal. Such abnormalities have been seen before in fermionic systems. Both natural and Dyson orbitals can be used to represent the density of the system. However, natural orbitals representation is demonstrated to be superior. With increasing boson numbers the difference between Dyson and natural orbitals becomes less prominent and it is expected to disappear in infinitely large systems of identical bosons.

\end{abstract}

\section{Introduction}
\label{intro}

Theoretical description of a system comprised of interacting particles often invokes a concept of single-particle motion in a self-consistent mean field. The idea of independent particles occupying single-particle orbitals finds wide applications in atomic, molecular, solid state and nuclear physics. It provides a simplified picture of a complicated problem making it more intuitive and manageable. 

The simplest way to construct single-particle orbitals is to use the  Hartre-Fock approximation to the many-body problem. Each such orbital is characterised by a single-particle energy and   wave function. The latter exponentially decreases  at large distances from the mean field with a rate determined by the single-particle energy. 
Alternatively, the orbitals and their occupation numbers could be extracted from a known many-body wave function by diagonalizing  the nonlocal one-body density matrix of the system. Such orbitals are called ``natural orbitals" and, unlike Hartree-Fock single-particle wave functions, they can change when the system undergoes transitions from one excited state to another.
%\color{red}
For any many-body system the sum of squared natural orbitals is equal to its density distribution,
however, one cannot disentangle individual components 
 from this sum through direct  observations.
%Natural orbitals are not observables but they  could be an efficient choice of single-particle basis states for constructing many-body wave functions in electronic, atomic and nuclear physics 

Natural orbitals  could be an efficient choice of single-particle basis states for constructing many-body wave functions in electronic, atomic and nuclear physics 
\color{black}
(see, for example, \cite{Dav72,Lew88,Fas22} and references therein). 
%They can also be used to gain information about momentum distirubion of many-body systems  constituents \cite{Lew88}.
 %\color{blue}
 For small systems comprised of identical particles the use of natural orbitals basis  could be more problematic   due to the necessity of dealing with the translation invariance of the system. \color{black} However, natural orbitals could be still constructed in this case.   An extreme example of natural orbitals in a three-body system could be found in \cite{Suz02} for a $3\alpha$  system representing the $^{12}$C nucleus, showing very  different orbital behaviour in the ground and excited states.

The naive single-particle picture of a many-body system implies that
removing one of its particles  would result in changed occupancies of the orbitals. However, the orbitals themselves change, both in the Hartee-Fock and natural orbital pictures. The probability of the physical process of removal depends on properties of both the  initial and final states of the system. It is determined by the overlap between the many-body wave functions of the initial and final states,  called Dyson orbital in atomic physics \cite{Ort20} and overlap functions (or  overlap integrals\footnote{%\color{red}
The ``overlap integrals'' are not  numbers but functions  because the integration is performed  not  over all the wave functions' arguments. Quite often the number of leftover variables is specified by using  terms like {\it one-nucleon overlap  integrals} or {\it two-nucleon overlap integrals} etc.}
\color{black} ) in nuclear physics \cite{Ber65}. An important property of these overlaps is that at large distances they decrease exponentially with the rate determined by the separation energy of the removed particle. The link between natural and Dyson orbitals have been discussed, for example, in \cite{Van93}. In particular, they should share the same asymptotic decrease and both can represent the matter density of the system.

%Since natural orbitals form a complete basis, they can be used for expanding the Dyson orbitals over them.

In general, it is difficult to achieve the correct asymptotic behaviour of Dyson orbitals using many-body wave functions expanded over some basis functions that decrease faster than the Dyson orbitals do (for example, the harmonic oscillator basis). However, if only those basis functions in the expansion are selected that contribute to long-range behaviour then convergence to the correct asymptotics is accelerated. This has been demonstrated in the Hypersphecial Cluster Model (HCM), originally introduced for fermions in \cite{Tim07} and reformulated in \cite{Tim23} for bosonic systems. The structure of hyperspherical cluster harmonic (HCH) basis functions resembles the one used in harmonic-oscillator-based microscopic cluster models \cite{Des12} but it has an advantage of being independent of the harmonic oscillator radius. The latter could be considered as a generator coordinate in the  hyperradial wave functions expansion so that solving the Schr\"odinger equation in hyperspherical coordinates should provide an optimal distribution over the harmonic oscillators with different radii. 

In \cite{Tim23}, the HCM Dyson orbitals were constructed for small helium drops containing from five  to ten  atoms. Interaction via a  soft two-body gaussian potential was considered there to concentrate on convergence of the asymptotic behaviour only. In constructing the Dyson orbitals, explicit information about the system with one   removed boson has been introduced. On the other hand, no explicit information about any subsystems is introduced in natural orbitals calculations. It is therefore interesting to investigate the relation of these two quantities, one of which does and another does not explicitly depend on the information about the subsystem with one missing boson. It is particularly interesting to see how this relation changes with the number of bosons in the system. 

 In this paper, the HCM has been extended to enable calculations of the translation-invariant nonlocal one-body density matrix. 
% \color{red} 
 This represents a novel application of the HCM towards a quantity that has not been studied for the few-body systems selected here. Importantly, this is   the first study of  natural orbitals that  explicitly accounts for translation invariance for light systems with more than four particles, specifically,  for small helium drops with less than 20 atoms.
 With the  helium-helium  interaction, \color{black} 
 employed earlier in \cite{Tim23}, the natural orbitals both in ground and first-excited states are constructed and their link with the Dyson orbitals is clarified. The densities of these drops are discussed in the context of both natural and Dyson orbitals. Sec. II introduces the translation-invariant definition of the nonlocal one-body density matrix and natural orbitals. Expressions for nonlocal density in the HCH basis are derived in Sec. 3. Numerical results for natural and Dyson orbitals are given in Sec. 4 and 5, respectively. Sec. 6 discusses their relation to matter density while Sec. 7 summarises and discusses the results obtained and draws conclusions. All necessary details of mathematical derivations are given in the Appendix.

\section{Density matrix and natural orbitals}

The natural orbitals $n(\ve{r})$ are the eigenfunctions of the nonlocal density matrix $D(\ve{r},\ve{r}')$:
\beq
\int d\ve{r}' D(\ve{r},\ve{r}') n(\ve{r}') = \lambda \,n(\ve{r}).
\eeqn{natorbdef}
For finite systems, composed of $A$ particles and described by the translation-invariant  wave function $\Psi$, the density matrix is defined in terms of the  positions  $\ve{r}$ and $\ve{r}'$ with respect to the centre of mass of %\color{red} 
the $A$-body system \color{black} as
\beq
D(\ve{r},\ve{r}') = \left(\frac{A-1}{A}\right)^{3/2}\la \Psi | \sum_{i=1}^A \delta(\ve{r} - (\ve{r}_i - \ve{R})) \delta(\ve{r'} - (\ve{r}'_i - \ve{R}')) 
| \Psi \ra,
\eeqn{nldens}
where $\ve{r}_i$ is the position of particle $i$ with respect to a chosen origin in the lab system and $\ve{R} = \sum_i \ve{r}_i/A$ is the centre of mass of the $A$-body system. The  factor $\left(\frac{A-1}{A}\right)^{3/2}$ guarantees that the   diagonal density $\rho(\ve{r},\ve{r})$, representing physical matter density distribution, is normalized to the number of particles $A$ in the system.

Only $0^+$ states are considered in this work, and for this case   the partial wave expansion of the nonlocal density is given by
\beq
D(\ve{r},\ve{r}') =  \frac{1}{rr'} \sum_{lm } D_l(r,r')  Y_{lm}(\hat{\ve{r}})  Y^*_{lm}(\hat{\ve{r}}').
\eeqn{2}
Assuming the partial wave decomposition of natural orbital $n(\ve{r})$ is
\beq
n(\ve{r}) =  \sum_{lm} \frac{ n_l(r)}{r} Y_{lm}(\hat{\ve{r}}),
\eeqn{nexpand}
one can represent the eigenvalue problem of Eq. (\ref{natorbdef}) in the form
\beq
\int_0^{\infty} dr' D_l(r,r')n_l^{(k)}(r') = \lambda_l^{(k)} n_l^{(k)}(r),
\eeqn{natorbdef_partial}
where $k$ marks the $k$-th eigenvalue. 
The eigenvalues $\lambda_l^{(k)}$ give occupation numbers of natural orbitals with orbital momentum of $l$ and the  sum over all $l$ is equal to $A$. The radial nonlocal density of the system is then
\beq
D_l(r,r') =  \sum_{k}\lambda_l^{(k)}n_l^{(k)}(r)n_l^{(k)}(r'),
\eeqn{partial_density}
while the diagonal density is
\beq
%\rho(r) = 
D(\ve{r}) \equiv D(\ve{r},\ve{r})
= 
%   Uncomment to see derivation
%
%\frac{1}{r^2} \sum_{lmk} \lambda_l^{(k)}\left(n_l^{(k)}(r) \right)^2  (lm l-m |0 0) (l 0 l | 0 0 ) (2l+1)/ {4\pi}\eol =
%\frac{1}{4\pi r^2} \sum_{lmk} \lambda_l^{(k)}\left(n_l^{(k)}(r) \right)^2 (-)^{l-m}(-)^l (2l+1) /(2l+1) \eol
\frac{1}{4\pi r^2} \sum_{lk} \lambda_l^{(k)}\left(n_l^{(k)}(r)  \right)^2.
\eeqn{diagonal_density}
%\color{red} 
Note that because only spin-zero  states are considered here this density is spherically-symmetrical.
\color{black}

\section{Nonlocal one-body density in the hyperspherical cluster model}

The translation-invariant one-body nonlocal density could be explicitly written in normalised Jacobi coordinates $\{\ve{\xi}_i\}\equiv \{\ve{\xi}_1,\ve{\xi}_2,\dots,\ve{\xi}_{A-1}\}$ defined as,
\beq
\ve{\xi}_{A-i}=\sqrt{\frac{i}{i+1}} \left(\frac{1}{i} \sum_{j=1}^i \ve{r}_j - \ve{r}_{i+1} \right),
\eeqn{Jacobi}
where $\ve{r}_ i$ is the  individual coordinate of the $i$-th particle. This definition uses $\ve{\xi}_{A-i}$ in the l.h.s. of (\ref{Jacobi}) instead of the standard choice  of $\ve{\xi}_i$  to simplify notations in the sections below. 
The nonlocal density definition is then
\beq
\rho(\ve{r},\ve{r}') 
=
A \int d\tau_{A-1} d\ve{\xi}_{1}d\ve{\xi}'_{1} \,\Psi^*(\ve{\xi}_{A-1},...,\ve{\xi}_{2},\ve{\xi}_{1})
 \eol
\times
\delta\left(\ve{r} - \alpha \ve{\xi}_{1}\right)\delta\left(\ve{r}' - \alpha \ve{\xi}'_{1}\right) \,\Psi(\ve{\xi}_{A-1},...,\ve{\xi}_{2},\ve{\xi}'_{1}),
\eeqn{nl_density_def}
%%%%%%%%%%%%%%
%%% Original version with more detials:
%%%%%%%%%%%%%%%%
%\beq
%\rho(\ve{r},\ve{r}') = \la \Psi | \sum_{i=1}^A \delta(\ve{r} - (\ve{r}_i - \ve{R})) \delta(\ve{r'} - (\ve{r}'_i - \ve{R}')) | \Psi \ra
%\eol
%=
%A \int d\ve{\xi}_1...d\ve{\xi}_{A-1} d\ve{\xi}'_1...d\ve{\xi}'_{A-1} \Psi^*(\ve{\xi}_1,...\ve{\xi}_{A-1})\Pi_{i=1}^{A-2}\delta(\ve{\xi}_i -\ve{\xi}'_i) \eol\times\delta\left(\ve{r} - \sqrt{\frac{A-1}{A} }\ve{\xi}_{A-1}\right)\delta\left(\ve{r}' - \sqrt{\frac{A-1}{A} }\ve{\xi}'_{A-1}\right) \,\Psi(\ve{\xi}'_1,...\ve{\xi}'_{A-1})
%\eol
%=
%A \int d\tau_{A-1} d\ve{\xi}_{A-1}d\ve{\xi}'_{A-1} \,\Psi^*(\ve{\xi}_1,...\ve{\xi}_{A-1})
 %\eol \times
%\delta\left(\ve{r} - \sqrt{\frac{A-1}{A} }\ve{\xi}_{A-1}\right)\delta\left(\ve{r}' - \sqrt{\frac{A-1}{A} }\ve{\xi}'_{A-1}\right) \,\Psi(\ve{\xi}'_1,...\ve{\xi}'_{A-1})
%\eeqn{1}
where $ d\tau_{A-1} =d\ve{\xi}_{A-1}...d\ve{\xi}_{2} $ and $\alpha = \sqrt{\frac{A-1}{A} }$. In fact, the choice of all Jacobi coordinates except $\ve{\xi}_1$ is not important. The latter  should be defined in terms of the position of the particle $A$ with respect to the centre-of-mass of $A-1$.
%$\ve{\xi}_{1} = \alpha\,(\ve{r}_A - \ve{R})$.

In this work the wave function $\Psi$ is expressed in hyperspherical coordinates $\ve{\rho} \equiv \{\rho,\hat{\rho} \}$ which are based on the canonical choice of Jacobi coordinates (\ref{Jacobi}). In these coordinates the hyperradius $\rho$ is defined as
%The Jacobi coordinates form the $(A-1)$-dimensional vector $\ve{\rho}$ with the length   given by the hyperradius $\rho$,
\beq
\rho^2 = \sum_{i=1}^{A-1} \ve{\xi}_i^2 =
%\sum_{i=1}^A \ve{r}_{ i}^2 -  \ve{R}^2  =
\frac{1}{A}\sum_{i < j}^A(\ve{r}_{i}-\ve{r}_{ j})^2.
\eeqn{f1}
%where $\ve{R} = (\sum_{i=1}^A \ve{r}_{i})/\sqrt{A}$ is the normalised  coordinate of the centre of mass. 
The hyperangles associated with $\ve{\xi_2}$ and $\ve{\xi}_1$ are chosen as $\{\theta_2,\hat{\xi_2},\theta_1,\hat{\xi}_1\}$ with
\beq
\xi_{1} &=& \rho \cos \theta_{1},
\eol
\xi_{2} &=& \rho \sin\theta_1 \cos \theta_2.
\eeqn{}
%and we also need the hyperradii in other-dimensional spaces $\rho_c$ and .....
The wave function $\Psi$ is expanded over the bosonic HCH basis, representing eigenstates of the angular part of the kinetic energy operator \cite{Tim23},
\beq
{\cal Y}_{\nu}(\hat{\ve{\rho}}) =  {\cal N}^{-1}_{\nu} {\cal S} \left[ Y_0(\hat{\rho}_c) \varphi^{(n)}_{0 \nu 0} (\theta_1, \hat{\xi}_1)\right],
\eeqn{HCH0}
where $Y_0(\hat{\rho}_c)$ is the lowest order hyperspherical harmonics (HH) for the $A-1$ core,  \beq
{\cal S} = \frac{1}{A^{1/2}} \left(1 + \sum_{i=1}^{A-1} P_{iA}\right)
\eeqn{S}
%\color{red} 
is the symmetrizer  that contains the operator $P_{iA}$
  permuting the $A$-th boson with bosons of the $A-1$  core. The definitions  of the hyperangular function $\varphi^{(n)}_{0 \nu 0}$,  that  enters (\ref{HCH0}), and the normalization coefficient ${\cal N}_{\nu}$, are given in the Appendix. \color{black} Retaining only the lowest order HH in (\ref{HCH0}) is justified by using a soft two-body potential in the calculations \cite{Tim23}. With this choice %for the $A-1$ core  
the set (\ref{HCH0}) is orthonormal, $\la {\cal Y}_{\nu}(\hat{\ve{\rho}}) \mid {\cal Y}_{\nu'}(\hat{\ve{\rho}}) \ra= \delta_{\nu,\nu'}$ and
the HCH expansion of $\Psi$  then reads
\beq
\Psi(\ve{\rho}) =
\rho^{-(n-1)/2} \sum_{\nu=0}^{\nu_{\max}} \chi_{\nu}(\rho) {\cal Y}_{\nu}(\hat{\rho}),
\eeqn{expansion}
where $n = 3A-3$ and $\chi_{\nu}(\rho)$ are the hyperradial functions found by solving a coupled system of differential equations %\color{red}
\beq
 \left(-\frac{d^2}{d\rho^2}+\frac{{\cal L}_{\nu}({\cal L}_{\nu}+1)}{\rho^2}
-\frac{2m}{\hbar^2} \left(E - V_{\nu\nu}(\rho)\right) \right)
\chi_{\nu}(\rho)
\eol
= -\frac{2m}{\hbar^2}
\sum_{\nu'\ne \nu} V_{\nu\nu'}(\rho) \chi_{\nu'}(\rho),
\eeqn{f5}
where  $m$ is 
the mass   of the helium atom,
${\cal L}_{\nu} = \nu + (n-3)/2$ is with the generalized angular momentum   
and $V_{\nu\nu'}(\rho)=
\la Y_{\nu}(\hat{\rho}) |  \sum_{i<j}V_{ij}(\ve{r}_i-\ve{r}_j)|
Y_{\nu'}(\hat{\rho}) \ra
$ are the hyperradial coupling potentials \cite{Tim23}. 
\color{black}

With   expansion ({\ref{expansion}), the nonlocal one-body density becomes
\beq
D(\ve{r},\ve{r}') = 
A \alpha^3 \sum_{\nu \nu'} 
\int_0^{\infty} d\rho_c \frac{\rho_c^{n-4}}{\rho^{n-1}}\chi_{\nu} (\rho)\chi_{\nu'}(\rho)
\,\,\,\,\,\,\,\,\,\,\,\,\,\,\,\,
\eol
\times
\int d {\ve{\xi}_1}d\ve{\xi}'_1\, {\la 
\cal Y}_{\nu}(\hat{\ve{\rho}})
\mid
\delta\left(\ve{r} - \alpha \ve{\xi}_{1}\right)\delta\left(\ve{r}' - \alpha \ve{\xi}'_{1}\right)
\mid
{\cal Y}_{\nu'}(\hat{\ve{\rho}}')
\ra ,
\eeqn{density_in_HH}
where $\rho_c= \sqrt{\rho_c^2 + \alpha^{-2} r^2}$ is the hyperradius of the $A-1$ core of the system and the integration in the matrix element ${\la 
\cal Y}_{\nu}(\hat{\ve{\rho}})
\mid
...
\mid
{\cal Y}_{\nu'}(\hat{\ve{\rho}}')
\ra $ is performed over the hyperangular variables $\hat{\rho}_c$ of  $A-1$. Using results (\ref{OBME_HCM}) from the Appendix    and
\beq
\int d {\ve{\xi}}d\ve{\xi}'\,
\varphi_{K' \nu l m}^{(n)*}(\theta,\hat{\xi})  
\, \delta\left(\ve{r} - \alpha \ve{\xi}\right)\delta\left(\ve{r}' - \alpha \ve{\xi}'\right)
\varphi_{K \nu' l m}^{(n)}(\theta',\hat{\xi}')
\eol
=
\alpha^6
\varphi_{K' \nu lm }^{(n)*}(\theta_r,\hat{r})  
\, 
\varphi_{K \nu' lm }^{(n)}(\theta'_r,\hat{r}'),
\,\,\,\,\,\,\,\,\,
\eeqn{}
where 
\beq
\cos \theta_r =  \frac{r}{\sqrt{\alpha^2 \rho_c^2 +   r^2}};
\,\,\,\,\,\,\,\,\,\,
\cos 2\theta_r = \frac{  r^2- \alpha^2 \rho_c^2}{\alpha^2 \rho_c^2 +  r^2};
\eol
\cos \theta'_r =   \frac{r'}{\sqrt{\alpha^2 \rho_c^2 +   r'^2}};
\,\,\,\,\,\,\,\,\,\,
\cos 2\theta'_r = \frac{  r'^2- \alpha^2\rho_c^2}{\alpha^2 \rho_c^2 + r'^2};
\eeqn{cos2th}
one obtains
\beq
D_l(r,r') 
=
\frac{\alpha^{-3/2}}{2l+1} \,
\sum_{\nu_2=0}^{\nu_{\max}-l} 
\int_0^{\infty}d\rho_c {\cal S}^{(b)}_{\nu_2l} (\rho_c,r){\cal S}^{(k)}_{\nu_2l} (\rho_c,r') 
\eeqn{rho_l}
where  
\beq
{\cal S}_{\nu_2l}^{(i)} (\rho_c,r) =\sqrt{ \frac{\rho_c^{n-4}}{\rho^{n-1}}}
\sum_{\nu=\nu_2+l}^{\nu_{\max}} \delta_{\nu_1,\nu-\nu_2-l} \delta_{K', 2\nu_2+l}
\,{\cal C}^{(i)}_{\nu\nu_1\nu_2l} \,  \chi_{\nu}(\rho)
 \phi_{K' \nu_1 l }^{(n)}(\theta_r)
\eeqn{calS}
with   
${\cal C}^{(i)}_{\nu\nu_1\nu_2l}$ given by (\ref{Cb_Ck}) from the Appendix.

\begin{figure}[t]
\vspace{0.5 cm}
\includegraphics[scale=0.31]{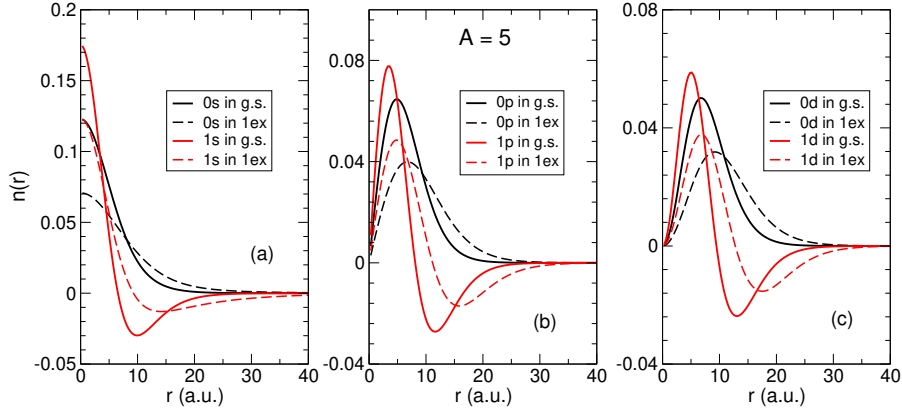}
\caption{The natural orbitals in ground (solid lines) and first excited (dashed lines) states of a five-boson system shown for $s$-, $p$- and $d$-waves in ($a$), ($b$) and ($c$), respectively. }
\label{fig:natorb.A=5}
\end{figure}

\begin{figure}[t]
\vspace{0.5 cm}
\includegraphics[scale=0.31]{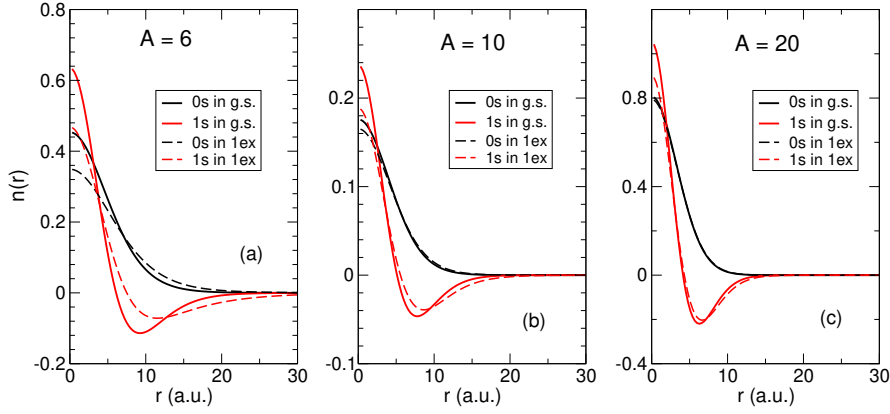}
\caption{The $s$-wave natural orbitals in ground (solid lines) and first excited (dashed lines) states of ($a$) six-, ($b$) ten- and ($c$) twenty-boson systems. }
\label{fig:natorb.61020}
\end{figure}

\section{Numerical results for natural orbitals}

The natural orbitals are constructed here for helium drops with  $A=5,6,8,10$ and 20 atoms using the soft two-body He-He Gaussian potential $V_{ij}(r) = V_0 \exp \left(-r^2/a_0^2\right)$ from \cite{Gat11} with $V_0 = -1.227$ K and $a_0 = 10.03$ a.u. The $\hbar^2/m=41.281 307$ (a.u.)$^2$K was chosen, as in previous works %\color{red} 
\cite{Tim23,Gat11,Tim12}\color{black} . With this choice of the two-body potential the convergence of the hyperspherical harmonics expansion accelerates with increasing $A$ \cite{Tim12} and the HCH  basis takes most of the responsibility for the convergence \cite{Tim23}.

The wave functions $\chi_{\nu}$ were found from solving a coupled system of differential equation using an  adaptation of the Lagrange-Laguerre method for hyperspherical coordinates as introduced in \cite{Tim17}.
More details on the precision of this solution and convergence can be found in \cite{Tim23,Tim17}.  The wave functions $\chi_{\nu}$ were then used to calculate  the ground- and first-excited states'  nonlocal one-body densities which then were diagonalised using the finite difference approach to get the natural orbitals and   their occupation numbers.

The first two $s$-, $p$- and $d$-wave natural orbitals  are shown in Fig. \ref{fig:natorb.A=5} for $A=5$. It is noticeable that they are very different in the ground and  first excited states. With increasing number of bosons this difference becomes smaller, which is illustrated in Fig. \ref{fig:natorb.61020}  showing the evolution of natural orbitals    with $A$  for 6, 10 and 20 bosons. For 20 bosons the difference between the $0s$ orbitals in  ground and first excited states is not noticeable anymore while 1$s$ orbitals still differs slightly. In all cases the natural orbitals in first excited states are more extended for all angular momenta, as expected.

\begin{figure}[b]
\vspace{0.5 cm}
\centering
\includegraphics[scale=0.33]{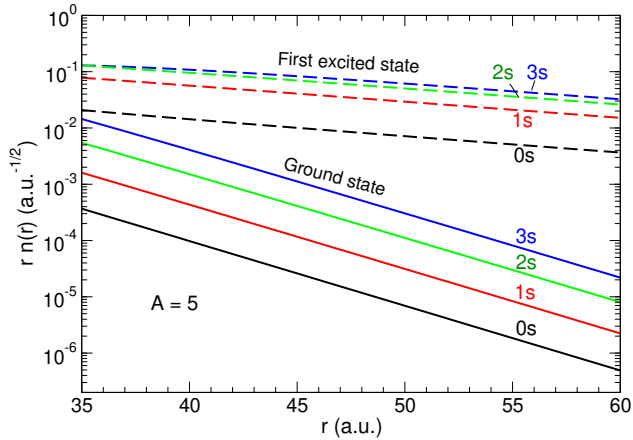}
\caption{Asymptotic behaviour  of four lowest $s$-wave natural orbitals $n(r)$, multiplied by $r$,  in the ground (solid lines) and first excited (dashed lines) states of a five-boson system. }
\label{fig:asym}
\end{figure}

Further examination of the natural orbitals revealed that  all orbitals with the same $l$, belonging to the same state of $A$, have exactly the same asymptotic behaviour  $ n_i(r) \sim \exp(-\kappa r)/r$ at $r \rightarrow \infty$, which is illustrated in Fig. \ref{fig:asym} for the case of the $s$-wave orbitals in $A=5$. Similar asymptotic behaviour was found for fermionic systems  in \cite{Van96}.
 It was found  here that for $0s$ orbitals the value of $\kappa$   equals exactly  to $ \sqrt{2(E_{A} - E_{A-1})/\mu}/\hbar$, where $\mu = (A-1)/A$, $E_A$ is the converged HCM energy of a chosen state in $A$ and   $ E_{A-1}$ is the $A-1$ ground state energy  obtained in the $\nu_{\max}=0$ version of the HCM. The orbitals with  higher node numbers reach the asymptotic form at larger $r$ but they all   eventually end up following the  same asymptotic decrease.
 
 It was pointed out in \cite{Van93} that natural orbitals  share the same asymptotic decrease with the Dyson orbitals. Since the natural orbitals are functions of the bosons' positions with respect to the centre of mass of $A$ while Dyson orbitals are defined in terms of their distances to the centre of mass of $A-1$, in small systems the  rate  $\kappa$ of the natural orbitals decrease  is related to the Dyson orbitals decrease $\kappa_D$ by $\kappa = \mu \kappa_D$. This indeed occurs in the present case, as confirmed by a comparison of the natural orbitals constructed here and the Dyson orbitals from \cite{Tim23}  at large $r$.  It is worth noting that the Dyson orbitals  in \cite{Tim23} explicitly contain the information about the $A-1$ system since they were constructed by overlaping the converged $A$-body solutions with the $\nu_{\max}=0$ ground state $A-1$-body wave functions. It is remarkable that the asymptotic decrease of the HCM natural orbitals turned out to be determined by the same separation energies  without introducing any input about the $A-1$ subsystem.

\begin{table*}
\caption{ Occupancies of natural orbitals in ground and first excited states of several helium drops.
}
      \begin{tabular}{p{0.8 cm}p{0.7 cm}p{1.1 cm}p{1.2 cm}p{1.2cm}p{1.2 cm}p{1.2cm}p{1.0 cm} }
\hline
\hline
 &  & 0s &1s & 2s &0p &1p &0d \\ 
\hline
A=5 & g.s. & 4.97  & 1.39e-2  & 1.15e-4 & 1.05e-2  & 1.05e-4  & 8.03e-4  \\
 & 1ex & 3.50 &       1.00 &  1.58e-2 &     0.381 &  1.00e-2   &   7.20e-2\\
 A=6 & g.s. &  5.98 & 9.54e-3  & 6.12e-5 & 7.93e-3  & 4.65e-5  &  3.25e-4\\
  & 1ex & 4.69 &    1.10&    6.64e-3 &  0.183&        1.49e-3   &   1.53e-2   \\
  A=8 & g.s. & 7.99 &        5.68e-3  & 3.89e-5 &
              5.12e-3 &  1.59e-5  & 9.05e-5 \\
           & 1ex & 6.82 &   1.12 &    2.61e-3&  6.07e-2    & 3.54e-4   &  1.63e-3 \\
A=10 & g.s. & 9.99&  3.96e-3 &  3.56e-5 &  3.72e-3 &  7.23e-6  &  3.69e-5 \\
   & 1ex &       8.87 &  1.10&   1.58e-3 & 
           3.00e-2 &  1.70e-4 & 3.88e-4 \\ 
A=20 & g.s. & 19.997   &     1.48e-3 &  3.64e-5 & 1.50e-3 &  6.47e-7   & 2.99e-6 \\
 & 1ex &  18.94 &       1.05 &     5.12e-4 &  5.42e-3  & 3.65e-5  & 1.16e-5 \\
\hline
\hline
\end{tabular}      
\end{table*}

The natural orbitals for  $A>5$  show similar asymptotic patterns. However, with increasing $A$ the difference between $\kappa$ in the ground and first excited state becomes smaller. 
%due to developing collapse of the bosonic systems with two-body interactions only. 
%Therefore, their asymptotic decays become more similar. 
It was also found that for a fixed state of $A$ the   $p$- and $d$-wave orbitals decrease faster  at large $r$ than the   $s$-wave orbitals do. The rate of asymptotic decrease of the $p$- and $d$-wave orbitals seems to be very similar.

The occupation numbers of the lowest natural orbitals are shown in Table I.  One can see that in the ground states almost all bosons occupy the $0s$ orbital even for the smallest system considered, with $A=5$. In the first excited state the occupancies of the $0s$ orbital are just below $A-1$, with $1.0-1.12$ particles occupying the $1s$ orbital, which is expected. Noticeable occupation of the $0p$ orbital is seen only in  $A=5$ and $A=6$ systems, while higher-$l$ orbitals remain practically unoccupied.

\section{Dyson orbitals and their link to natural orbitals}

\begin{table*}
\caption{ Expansion coefficient of Dyson  over  natural orbitals for several helium drops. The overlaps marked by $^*$ are expanded over natural orbitals of the first excited states while for in all other cases the ground-state natural orbitals basis was used.
}
      \begin{tabular}{p{2.6 cm}p{1.0 cm}p{1.1 cm}p{1.1cm}p{1.1cm}p{1.1cm}p{1.1cm}}
\hline
\hline
Dyson overlap  & 0s & 1s & 2s & 3s & 4s & 5s \\ 
\hline
$\la 5({\rm g.s.}) \mid 4({\rm g.s.})\ra$ & 1.546    &    1.98e-2  & 5.93e-4   & 2.60e-5 & 1.88e-6  &  2.47e-7\\
$\la 5({\rm 1ex})\mid 4({\rm g.s.})\ra^*$ &
-0.533 &       0.621 &    -1.09e-2  &-3.11e-2&
-4.20e-2 &  4.88e-2  %-4.7445297756228536E-002
 \\
$\la 5({\rm g.s.})\mid 4({\rm 1ex})\ra$ &
-0.239&    3.65e-2 &  3.02e-3  &  1.59e-4   & 2.22e-5  & 8.07e-6 
\\
$\la 5({\rm 1ex})\mid 4({\rm 1ex})\ra$ &
0.810  &      0.211&  8.51e-2 & 2.77e-2 &  1.02e-2  & 4.35e-3 \\
\hline
$\la 6({\rm g.s.})\mid 5({\rm g.s.})\ra$ &
-1.822    &   -1.42e-2 & -3.76e-4  & 1.75e-5  & -8.84e-7 & -1.36e-7  % 5.0645056017182291E-007
\\
$\la 6({\rm 1ex})\mid 5({\rm g.s.})\ra^*$ &
-0.433   &   -0.759    &   -7.78e-3 &   1.76e-3  & -3.38e-3 &  4.85e-3 % & -7.08e-003
\\
$\la 6({\rm g.s.})\mid 5({\rm 1ex})\ra$ &
 -0.311 &      5.51e-2  & 1.95e-3  & -7.47e-5   & -1.08e-5  &  2.53e-6\\
$\la 6({\rm 1ex})\mid 5({\rm 1ex})\ra$ &
  1.374 &     0.160 &    5.21e-2  & 1.00e-2  & -2.95e-3 &  1.05e-3 \\
\hline
$\la 8({\rm g.s.})\mid 7({\rm g.s.})\ra$ & -2.283   &    -9.08e-3 &  3.27e-4 & -1.22e-5 &  -4.48e-7  & 1.11e-8 
\\
$\la 8({\rm 1ex})\mid 7({\rm g.s.})\ra^*$ &
-0.348 &     0.846 &      4.87e-3   & 1.30e-4 &  2.62e-5  &  3.15e-6 % -2.4968847803849564E-004 
\\
$\la 8({\rm g.s.})\mid 7({\rm 1ex})\ra$ &
-0.319 &   5.50e-2 &  -4.93e-4  & -7.51e-5  & -1.51e-5 & -2.85e-7 \\
$\la 8({\rm 1ex})\mid 7({\rm 1ex})\ra$  &
 1.977&  7.99e-2 & -2.73e-2 &  2.08e-3 &  -9.47e-4 &  2.67e-4 %  7.6145479544676304E-005 
 \\
\hline
$\la 10({\rm g.s.})\mid 9({\rm g.s.})\ra$ &
-2.671&   -6.64e-3 & -3.49e-4   & 8.62e-6  & -2.40e-7 & -6.48e-9 
\\
$\la 10({\rm 1ex})\mid 9({\rm g.s.})\ra^*$ & -0.316&    -0.883&     3.64e-3  & -2.86e-5  & -3.65e-5  & -6.41e-7 % &  1.4897442360251300E-007 
\\
$\la 10({\rm g.s.})\mid 9({\rm 1ex})\ra$ &
-0.317 &       5.01e-2  & -5.46e-5 &  -2.31e-5  & -1.18e-5  & -3.85e-7\\
$\la 10({\rm 1ex})\mid 9({\rm 1ex})\ra$ &
 -2.418  &  -4.33e-2 &  1.84e-2  & -6.11e-4  & 5.96e-4 & -1.49e-4  \\
\hline
$\la 20({\rm g.s.})\mid 19({\rm g.s.})\ra$&
4.116&   2.75e-3 &  4.17e-4   & 1.59e-7 &  9.95e-7  & 1.43e-6 \\
$\la 20({\rm 1ex})\mid 19({\rm g.s.})\ra^*$ &-0.276&  -0.944 &       -1.91e-3   & 9.23e-7 & -1.06e-7 & -1.19e-6 
\\
$\la 20({\rm g.s.})\mid 19({\rm 1ex})\ra$ &
0.308 &      -3.47e-2 & 2.14e-4  & -2.01e-6  & -9.89e-6  & -1.32e-5 \\
$\la 20({\rm 1ex})\mid 19({\rm 1ex})\ra$ &
  3.952 &    -6.66e-3  & 6.90e-3  & 1.02e-5   &2.61e-5  & 5.04e-6 \\
\hline
\hline
\end{tabular}      
\end{table*}

The Dyson orbitals $I(\ve{r})$ are constructed by calculating the overlap integral $\la \Psi_{A-1} | \Psi_A\ra$ between  the wave functions $\Psi_A$ and $\Psi_{A-1}$ of the $A$- and $A-1$-body systems \cite{Ort20,Ber65}. An additional factor of $\sqrt{A}$ is normally introduced into the definition. The Dyson orbitals (or overlap functions) satisfy a Schr\"odinger-like inhomogeneous equation with the source term $\la\Psi_{A-1} \mid \sum_i V_{Ai} \mid \psi_A\ra$ vanishing at large separations between  $A-1$ and removed boson \cite{Tim98}. This dictates the Dyson orbitals' asymptotic decrease in the form of $\exp(-\kappa_D r)/r$, where $r$ is the position of the $A$'s boson with respect to the centre of mass of $A-1$ and $k_D = \sqrt{2\mu (E_A-E_{A-1})}/\hbar$.
For $0^+$ states in $A$ and $A-1$ the partial wave expansion of $I(\ve{r})$  has a simple form, $I(\ve{r})= I(r)Y_{00}(\hat{r})$. The norm of the radial orbital, called probability factors or pole strengths
in atomic physics and spectroscopic factor in nuclear physics, is
\beq
S = \int_0^{\infty} dr r^2 I^2(r).
\eeqn{SF}
Each radial Dyson orbital $I(r)$ could be expanded onto the natural orbital basis as
\beq
I(r) = \sum_i c_i \frac{n_i(r/\mu )}{r/\mu  }
\eeqn{Ir_expansion}
with expansion coefficients
\beq
c_i = \int_0^{\infty} dr \, r n_i(r) I( \mu r).
\eeqn{coef_ci}
The reason why the factor $1/\mu $  is introduced in the argument of $n(r)$ in (\ref{Ir_expansion}) is due to the 
 different definitions of $r$ in Dyson and natural orbitals, which is important for small systems but becomes less so for large $A$. The $r/\mu$ choice in $n_i$ provides a faster convergence of the expansion (\ref{Ir_expansion}) in the asymptotic region due to $rI(r)$ and $n_i(r/\mu)$ sharing the same rate of asymptotic decay. For a chosen Dyson orbital the sum of the expansion coefficients satisfies
\beq
\mu^3 \sum_i c_i^2 = S. 
\eeqn{sumci}

\begin{figure}[t]
\vspace{0.5 cm}
\centering
\includegraphics[scale=0.38]{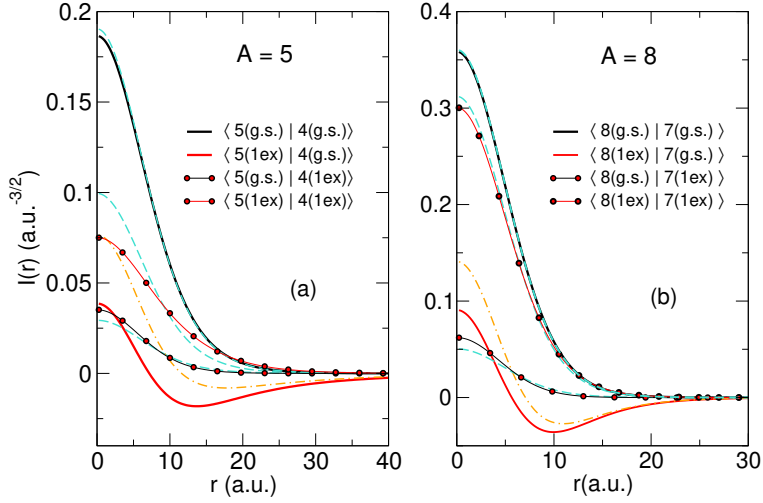}
\caption{Dyson orbitals (solid lines) in comparison to the leading term of their expansion over $s$-wave natural orbitals for five- ($a$) and eight-boson ($b$) systems. Dashed and dash-dotted lines correspond to the $0s$ and $1s$ natural orbitals, respectively.  }
\label{fig:Dysorb}
\end{figure}

The expansion coefficients for Dyson orbitals considered previously   in \cite{Tim23} and for $A=20$, are presented in Table II.  The natural orbital basis from ground-state density matrix diagonalization was used to determine $c_i$, except for the case of 
$\la A({\rm 1ex}) | A-1({\rm g.s.})\ra$, where natural orbitals were taken from the first excited state of $A$ to accelerate the expansion convergence (they are marked by $^*$ in the Table). In  Fig. \ref{fig:Dysorb},  the Dyson orbitals  are plotted for $A=5$ and 8 in comparison to the leading term of the expansion (\ref{Ir_expansion}). In all cases the Dyson overlaps $\la A({\rm g.s.}) | A-1({\rm g.s.})\ra$ are almost entirely represented by the 0$s$ natural orbitals  as expansion coefficients for higher orbitals are very small. 
%In  Fig. \ref{fig:Dysorb},  the Dyson orbitals  are plotted for $A=5$ and 8 in comparison to the leading term of the expansion (\ref{Ir_expansion}). 
The Dyson overlaps of the initial ground state with excited final states are also dominated by the $0s$ contribution, but contribution from the $1s$ state is still visible for the two cases shown in Fig. \ref{fig:Dysorb}.
As for the Dyson overlaps $\la A({\rm 1ex}) | A-1({\rm g.s.})\ra$, they turn out to be a mixture of the $0s$ and $1s$ natural orbitals, with $1s$ contribution becoming more prominent with increasing $A$. 

%For the lightest system considered here, $A=5$, the admixture of the $0s$ is strong, with some additional contribution from higher orbitals. This should be the consequence of the difference in the centre-of-mass positions of $A-1=4$ and $A=5$.

\begin{figure}[t]
\vspace{0.5 cm}
\centering
\includegraphics[scale=0.38]{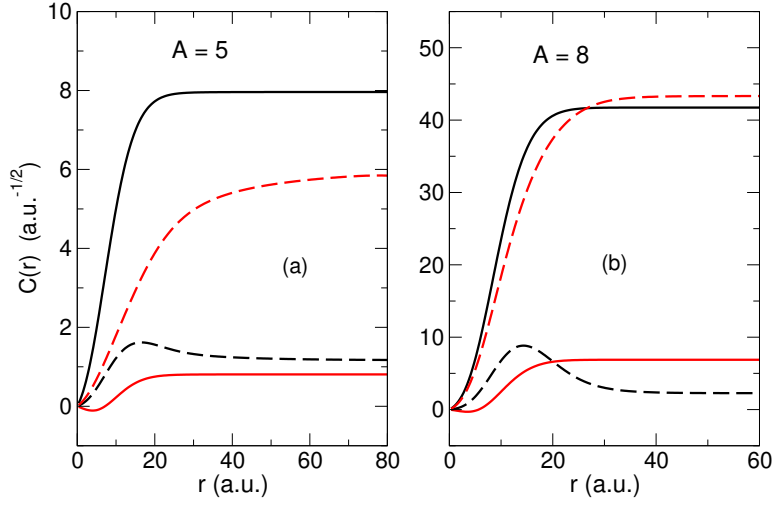}
\caption{Ratio $rI(r)/\exp(-\kappa_D r)$ of the Dyson orbital to their asymptotic forms for five- ($a$) and eight-boson ($b$) systems. Solid lines correspond to the final ground states while dashed lines correspond to first excited final states. Black and red curves correspond to the initial ground and first excited states, respectively.}
\label{fig:Dysorb_asym}
\end{figure}

The asymptotic behaviour of Dyson orbitals $\la A | A-1({\rm g.s.})\ra$  has been investigated in \cite{Tim23}. It converges to $\exp(-\kappa_D r)/r$ after $r \sim 25$ a.u., which is illustrated again in Fig. \ref{fig:Dysorb_asym}, where  the ratio $C(r) = rI(r)/\exp(-\kappa_D r)$ is plotted for $A=5$ and $A=8$. However, one can also see from this figure that the convergence pattern for $\la A| A-1({\rm 1ex})\ra$ is different, being  significantly  slower than in the case of the final ground states. The final shallow  $A-1$ excites states  could be considered as consisting of a halo boson moving around the $A-2$ core 
%, if halo  is associated with a considerable probability of staying outside this core 
\cite{Tim23} with significant spatial extension of the wave function $\Psi_{A-1}({\rm 1ex})$. Therefore, the source term $\la\Psi_{A-1}({\rm 1ex}) \mid \sum_i V_{Ai} \mid \psi_A\ra$, which governs the asymptotic decay of $I(r)$, should  be more  spatially extended than the one with $\Psi_{A-1}(g.s.)$, and this could be the origin of pre-asymptotic abnormalities in Dyson orbitals for excited final states.
%Removing one boson from the $A-1$ core leaves the final system in a state formed by a halo boson around the $A-2$ core. Then the $V_{A,A-1}$ interaction in the source term $\la\Psi_{A-1} \mid \sum_i V_{Ai} \mid \psi_A\ra$ can have a larger range due to the small  binding in the first excited state of $A-1$ thus causing a slow convergence of the Dyson orbitals to  their asymptotic form. %%Physically, this means that formation of the final excited state could happen by removing a boson from a correlated pair. 
%The halo  boson can still be in the vicinity of the $A$-th boson in $A$, even at large $r$, and their interaction can produce irregularity in the asymptotic behaviour of Dyson orbitals. 
The possibility of pre-asymptotic abnormalities was first pointed out for a special class of weakly-bound nuclei in \cite{Tim03}, where they were associated with triangle Feynman diagram describing nuclear virtual decay, and   then it was confirmed by direct three-body calculations in \cite{Tim08a,Tim08b}. In the particular case of $\la A=8({\rm 1ex}) \mid A=7({\rm 1ex}) \ra$ the ratio $C(r)$ looks similar to those described in these works. However the  $\la A=5({\rm 1ex}) \mid A=4({\rm 1ex}) \ra$ overlap has not reached its asymptotic form even at $r \sim 80$ a.u. despite showing a trend to convergence. The $\la A({\rm g.s.}) | A-1({\rm 1ex})\ra$ overlaps seem to achieve the asymptotic form faster, both for $A=5$ and $A=8$. But their behaviour at $10 \lesssim r \lesssim 25$ a.u. - the region within  the range of two-body potential   outside their cores - is very different from that observed in the case  of the ground state of $A-1$. The Dyson overlaps are significantly larger than their true asymptotics would suggest. The reason for this should be investigated further.

\section{Density}

\begin{figure}[t]
\vspace{0.5 cm}
\includegraphics[scale=0.32]{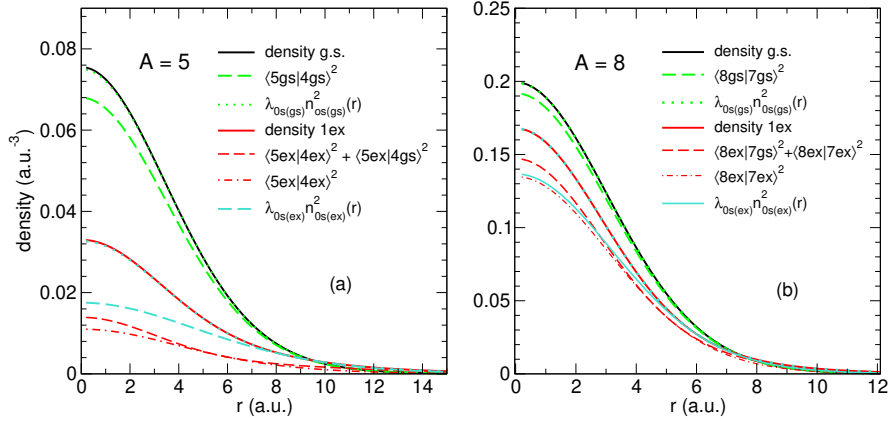}
\caption{Exact density of the ground and first excited state  of the $A=5$ ($a$) and $A=8$ ($b$) systems, shown by solid lines. Contribution   of the $0s$ orbitals to the excited state is shown by turquoise dashed lines.  The contribution of the Dyson orbitals $\la A({\rm g.s.)} ) \mid A-1({\rm g.s.})\ra$ to the ground state densities is represented by green dashed lines. The densities from $\la A({\rm ex)} ) \mid A-1({\rm ex})\ra$  are  shown by dashed-dotted red lines and their sum with $\la A({\rm ex)} ) \mid A-1({\rm g.s.})\ra$ is represented by red dashed lines. }
\label{fig:density}
\end{figure}

The density of the system has a simple representation in terms of natural orbitals as given by Eq. (\ref{diagonal_density}). Examination of the orbital occupancies from Table I suggests that  the ground-state densities of helium drops interacting by an attractive soft potential are given by the $0s$ natural orbitals only. Indeed, the $0s$ contribution, shown  in Fig. \ref{fig:density} for $A=5$ and $A=8$, cannot be distinguished from  exact density calculated from equations (\ref{rho_l}) and (\ref{calS}). Since  excited-state occupancies differ from the ground-state ones  approximately by one boson only, the $0s$ natural orbital contribution should also dominate the densities of excited states. However,
this contribution, shown in Fig. \ref{fig:density} by dashed turquoise lines,  is still not sufficient both for the $A=5$ and $A=8$ cases. Adding the $1s$ orbital contribution makes the $0s$+$1s$ density indistinguishable from its exact form. The excited state density spreads out further than the ground state one does and, therefore, the  centre of the helium drop is less dense. This difference gets smaller with increasing $A$ and one can expect it to disappear for $A \rightarrow \infty$.

The matter density could be represented in terms of Dyson orbitals $I_i$, associated  with overlaps with the wave function of  $A-1$ in the  state $i$ as \cite{Tim98}
\beq
D(r) = \frac{\mu^{-3} }{4\pi} \sum_{i } I^2_i\left(\frac{r}{\mu}\right).
\eeqn{density_from_OI}
 At $r \rightarrow \infty$ the asymptotic behaviour of the density is determined by the asymptotic behaviour of the ground-state Dyson orbital, since all the rest decrease faster. The contribution from this orbital dominates  in the ground state densities, which is seen from Fig. \ref{fig:density}, but it offers a poorer approximation for the density than the $0s$ natural orbital term does.  However, this  contribution  becomes   more important with increasing $A$.  The approximation of the excited-state density by one or two lowest Dyson orbital is even worse, although it is perfect in the asymptotic region (not shown in the figure). To get anywhere near the exact density the  sum of a large number of Dyson orbitals, corresponding to high excitations of $A-1$, is required, which at first sight seems inconsistent with relatively small norms of these orbitals (see Table 4 in \cite{Tim23} showing spectroscopic factors). However, such norms are not sensitive to the $I_i(r)$ behaviour at $r \approx 0$ while for densities it is crucial. A brief investigation  was carried out for $A=5$ to check  if overlap, for example,  with seventh excited state of $A=4$ gives a noticeable contribution to the excited-state density. It turned out that it does. With increasing $A$ the contributions from highly-excited states become less important and, as in the case of ground states, one can expect that for $A \rightarrow \infty$ only one Dyson orbital will contribute to the excited-state density.

\section{Discussion and conclusions}

The natural orbitals in small helium drops, constructed in HCM for soft gaussian two-body interactions, strongly depend on the boson number in the system and on its state. Only the $0s$ orbit shows significant occupation in the ground  states for all the cases considered, and this occupation increases with the boson number. The $0s$ occupancy in the   first excited state tends to $A-1$ with occupancy of the $1s$ orbit being slightly larger than one. With increasing $A$ the difference between the natural orbitals in ground and excited states decreases. %\color{blue} 
One reason for constructing natural orbitals is their potential to be used as a basis for configuration-interaction type of many-body calculations. It could be difficult to employ the present orbitals for such purposes since they are defined with respect to the centre-of-mass of small systems. Any calculations with these translation-invariant orbitals would involve many-body matrix elements, which are  difficult to handle. However, in the limit of $A \rightarrow \infty$, where the centre-of-mass motion becomes unimportant, this basis could find its applications. \color{black} Most likely, the $0s$ orbitals will be 100 per cent occupied in the ground state, representing a condensed bosonic system.

Removal of one boson from a system  is characterized by Dyson orbitals that capture information about both its initial and  final states. The Dyson orbitals could be readily expanded over the natural orbitals' basis. The $0s$ natural orbital dominates the $\la A({\rm g.s.}) \mid A-1 ({\rm g.s.}) \ra$ overlaps, almost exhausting its strength, it  also gives the main contribution to the Dyson orbitals 
 $\la A({\rm g.s.}) \mid A-1 ({\rm 1ex}) \ra$, although the $1s$ contribution  is also noticeable. The overlap $\la A({\rm 1ex}) \mid A-1 ({\rm 1ex}) \ra$ needs more natural orbitals in its expansion. Once again, with increasing $A$ only one natural orbital will remain in the  Dyson orbitals' expansions over natural orbitals.

 The $\la A  \mid A-1 ({\rm g.s.}) \ra$ overlaps and $0s$ natural orbitals share  the same asymptotic decrease (after a trivial rescaling of their arguments). This decrease is achieved in the region where the two-body interaction is vanishingly small. Unlike natural orbitals, which all share a common rate of decrease, the $\la A  \mid A-1 ({\rm 1ex}) \ra$ Dyson  overlaps with excited final states decrease faster according to the  (larger) separation energies needed to populate them. In addition, they show abnormalities  in achieving the asymptotic behaviour, associated with the interaction of the weekly-bound boson in $A-1$ with the boson undergoing removal from $A$. These abnormalities are significant and further investigation of their nature is needed. In particular, 
 a  clarification is needed on how extending the HCM to include more core states could affect these abnormalities.

 The density of the system can be expressed both in natural and Dyson orbitals. The examples considered here suggest that natural orbital expansion has an advantage. This advantage should disappear in infinitely large systems. %\color{red} 
 The Dyson orbitals could be readily used for calculating transition density distributions and  transition strengths for various one-body operators that determine observables, for example, electromagnetic ones. The  transition 
 densities  are just the products of the Dyson orbitals  corresponding  to the two states involved in the transition  summed over all configurations with one particle removed. The transition strengths  involve integration of the transition densities with  the radial form of the operator causing the  transition.
 The usage  of natural orbitals for the same purposes  needs a more detailed investigation. \color{black}

 One should keep in mind that all results obtained here are based on a soft gaussian two-body potential, which leads to %\color{red} 
 unphysical   growth of the  binding energy per atom  as $A \rightarrow \infty$  resulting in   the collapse of large bosonic systems. 
 To prevent this collapse,  an effective repulsive three-body force could be introduced,  and a simple   one-gaussian hypercentral repulsion   is  sufficient to  achieve saturation in the systems with large helium atoms
 \cite{Gat11,Kie17}. \color{black} However, this noticeably deteriorates the convergence of the HH expansion \cite{Tim15} and it is not yet clear how much of the convergence could be recovered with the HCM version with only one core state, employed here. Future work will  %\color{red} 
 include the three-body repulsion into the HCM scheme and  to clarify if the 
 present conclusions, regarding both  natural and Dyson orbitals, remain valid. 

 It should be pointed out that if a few-body system interacts with an environment, such as in the case of helium  clusters trapped in tungsten \cite{Wan20}, then the interaction with  the environment will change both the natural and Dyson orbitals. The HCM could still be adapted for their study in this case. Both orbitals will be expressed in different  coordinates, chosen  to suit best  the trap  geometry  under consideration.   The  HCM  would be particularly efficient for harmonic  traps,  similar  to those considered in \cite{Tol11}.

 Finally, the mathematical  aspects  of this work will  be very  useful  for further  developments  of  the HCM towards scattering and reaction calculations in atomic, molecular  and  nuclear physics. Such calculations require knowledge of the target densities to generate folding potentials.  Also, the Dyson orbitals,  as well as the hyperangular transformation coefficients from the Appendix,  are indispensable for including various exchange processes in the description of such processes.

\color{black}
\renewcommand{\thesubsection}{\Alph{subsection}}
\renewcommand{\theequation}{\Alph{subsection}.\arabic{equation}}
\setcounter{equation}{0}

\section*{Appendix}

\subsection{Hyperangular functions}
\label{sec:A}

The hyperangular function in the $n$-dimensional space is given by
\beq
\varphi_{K_c \nu lm}^{(n)} (\theta,  \hat{{\xi}})&=&  \phi_{K_c \nu lm}^{(n)} (\theta)
Y_{lm}(\hat{ { \xi}}),
\eol
\phi_{K_c \nu lm}^{(n)} (\theta)&
=&
N^{(n)}_{ K_c\nu l} \,
\cos^l\theta  \sin^{K_c} \theta \,
P_{\nu}^{K_c+\frac{n-5}{2},l+\frac{1}{2}}(\cos 2\theta),
\eeqn{varpi}
where $P$ is the Jacobi polynomial
  and
the normalization is given by
\beq
N_{nK_cl}^{(n)} = % \left(
\sqrt{
\frac{2\nu!\,(2\nu+K_c+l+(n-2)/2)\Gamma(\nu+K_c+l+(n-2)/2)}
{ \Gamma(\nu+l+3/2) \,\Gamma(\nu+K_c+(n-3)/2)}
}
%\right)^{1/2} 
.\,\,\,\,\,\,\,\,\,\,\,\,
\eeqn{nkkl}
In (\ref{varpi})   the $\hat{ { \xi}}$ are the angular variables of the Jacobi vector $\ve{\xi}$ and $\theta$ is the hyperangle which is related to hyperradii $\rho$ and $\rho_c$ in $n$- and $n-3$-dimensional spaces by
 \beq
 \sin \theta = \rho_c/\rho ,  \,\,\,\,\,\,\,\,\,\,\,
\cos \theta =  \xi/\rho,  \,\,\,\,\,\,\,\,\,\,\,
\cos 2\theta =  2\xi^2/\rho^2-1.
\eeqn{sincos}

\subsection{One-body matrix elements in hyperspherical cluster basis}

To construct the nonlocal density matrix in the HCM one first needs to evaluate the matrix elements
\beq
{\la 
\cal Y}_{\nu}(\hat{\ve{\rho}})
\mid
{\cal O}(\ve{r},\ve{r}',\ve{\xi}_{A-1},\ve{\xi}'_{A-1}) 
\mid
{\cal Y}_{\nu'}(\hat{\ve{\rho}}')
\ra,
\eeqn{OBME}
for an arbitrary single-particle operator ${\cal O}(\ve{r},\ve{r}',\ve{\xi}_{A-1},\ve{\xi}'_{A-1})% = \delta\left(\ve{r} - \alpha \ve{\xi}_{A-1}\right)\delta\left(\ve{r}' - \alpha \ve{\xi}'_{A-1}\right)
$, where the integration is performed  over the hyperangles $d\hat{\ve{\rho}}_c$ of the core $A-1$.
This will include evaluating several terms arising due to the symmetrization:
\beq
{\la 
\cal Y}_{\nu}(\hat{\ve{\rho}})
\mid
{\cal O}
\mid
{\cal Y}_{\nu'}(\hat{\ve{\rho}}')
\ra 
= ({\cal N}_{\nu} {\cal N}_{\nu'})^{-1/2}
%\eol
%\times
\left[
\la 
  Y_0(1,...,A-1)\varphi_{\nu}(A)
\mid
{\cal O}
\mid
{\cal Y}_{\nu'}(\hat{\ve{\rho}}')
%{\cal S} [Y_0(1,...,A-1)\varphi_{\nu'}(A')]
\ra \right.
\eol
+ \left.
 (A-1)\la Y_0(1,...,A-2,A )\varphi_{\nu}(A-1)
\mid
{\cal O}
\mid
{\cal Y}_{\nu'}(\hat{\ve{\rho}}')
%{\cal S} \left[Y_0(1,...,A-1)\varphi_{\nu'}(A')\right]
\ra
\right].
\,\,\,\,\,\,\,\,\,\,\,\,\,\,\,\,\,\,\,\,\,\,\,\,\,\,\,\,\,\,
\eeqn{MEcore}
Here $\varphi_\nu$ stands for a short notation for $\varphi_{0\nu0}$. The $\varphi_\nu(A)$ depends on $\hat{\xi}_1 $ and $\theta_1$ while $\varphi_\nu(A-1)$ depends on $\hat{\xi}''_1 $ and $\theta''_1$, see Fig. \ref{fig:coord-transform}. 
 The core harmonics in (\ref{MEcore}) can be represented as
\beq
Y_0(1,...,A-1 ) &=& Y_0(1,...,A-2) \varphi_{000}^{(n-3)}(\theta_2,\hat{\xi}_2),
\eol
Y_0(1,...,A-2,A ) &=& Y_0(1,...,A-2) \varphi_{000}^{(n-3)}(\theta''_2,\hat{\xi}''_2).
\eeqn{HHcore}
Then the product $\varphi_{000}^{(n-3)}(\theta''_2,\hat{\xi}''_2)\varphi_{0\nu 0}^{(n)}(\theta''_1,\hat{\xi}''_1)$ in the second term of (\ref{MEcore}) could be re-expressed in the coordinates $\{\hat{\xi}_2,\theta_2,\hat{\xi}_1 , \theta_1\}$, related  to $\{\hat{\xi}''_2,\theta''_2,\hat{\xi}''_1 , \theta''_1\}$  by the orthogonal transformation 
\beq
\ve{\xi}_1 &=& \gamma_1 \ve{\xi}''_1 +\gamma_2 \ve{\xi}''_2 \eol
\ve{\xi}_2 &=& -\gamma_2 \ve{\xi}''_1 +\gamma_1 \ve{\xi}''_2   
\eeqn{c-transform}
with $\gamma_1^2 +  \gamma_2^2 = 1$,  using  $\gamma_1 = -1/(A-1)$ and $\gamma_2 = \sqrt{A(A-2)}/(A-1)$. This is achieved by  using   relation 
\beq
\varphi_{000}^{(n-3)}(\theta''_2,\hat{\xi}''_2)\varphi_{0\nu 0}^{(n)}(\theta''_1,\hat{\xi}''_1)= \sum_{\nu_1 \nu_2 l} T^{(n)} _{\nu,\nu_1\nu_2l}\left[\varphi_{0\nu_2 l }^{(n-3)}(\theta_2,\hat{\xi}_2)\times\varphi_{K' \nu_1 l }^{(n)}(\theta_1,\hat{\xi}_1)\right]_{00},
\eol
\eeqn{Transform1}
where $\nu_1+\nu_2+l=\nu$, with the coefficients  $T^{(n)} _{\nu,\nu_1\nu_2l}$ derived in \cite{Tim23}. 
%%%%%%%%%%%%%%%%%%%%%%%%%%%%%%%%%%%%%%%%%
%%
%%   Uncomment for more detail
%%
%%%%%%%%%%%%%%%%%%%%%%%%%%%%%%%%%%%%%%%%%
%Then
%\beq
% Y_0(1,...,A-1)\varphi_{\nu}(A)+ (A-1)\la Y_0(1,...,A-2,A )\varphi_{\nu}(A-1)=\,\,\,\,\,\,\,\,\,\,\,\,\,\,\,\,\,\,\,\,
%\eol
% \sum_{\nu_1 +\nu_2 +l=\nu} \left(\delta_{\nu_2,0}\delta_{l,0} + (A-1)T^{(n)} _{\nu,\nu_1\nu_2l}\right)\left[\varphi_{0\nu_2 l }^{(n-3)}(\theta_2,\hat{\xi}_2)\times \varphi_{K' \nu_1 l }^{(n)}(\theta_1,\hat{\xi}_1)\right]_{00}
% \,\,\,\,\,\,
%\eeqn{}
The ket-vector in (\ref{MEcore}) could be is represented  in the following way %according to coordinate transformations shown in Fig.x:
\beq
A^{1/2} {\cal S} [Y_0(1,...,A-1) \varphi_{\nu'}(A)]
=Y_0(1,...,A-2)\, \varphi^{(n-3)}_{000}(A-1) \,\varphi^{(n)}_{\nu'}(A')
\eol
+
Y_0(1,...,A-2) \,\varphi^{(n-3)}_{000}(A') \,\varphi^{(n)}_{\nu'}(A-1)
\eol
+
\sum_{i=2}^{A-1}Y_0(...,A-i-1,...) \,\varphi^{(n-6)}_{000}(A-1) 
%\,\underbrace{
\varphi^{(n-3)}_{000}(A') \,\varphi^{(n)}_{\nu'}(A-i)
%}_{\rm transformation \,(1)}
. \,\,\,\,\,\,\,\,\,\,\,\,
%%%%%%%%%%%%%%%%%%%%%%%%%%%%%%%%%%%%%%%%%%%%%%%%%%%
%%
%%   Uncomment for more detail
%%
%%%%%%%%%%%%%%%%%%%%%%%%%%%%%%%%%%%%%%%%%%%%%%%
%\eol
%=
%Y_0(1,...,A-2)\sum_{\nu'_1 \nu'_2 l'} \left( \delta_{\nu'_20}\delta_{l'0}+T^{(n)} _{\nu',\nu'_1\nu'_2l'}\right) \left[\varphi_{0\nu'_2 l' }^{(n-3)}(\theta_2,\hat{\xi}_2)\varphi_{K' \nu'_1 l' }^{(n)}(\theta'_1,\hat{\xi}'_1)\right]_{00}
%\eol
%+
%\sum_{i=2}^{A-1}Y_0(...,A-i-1,...) \,
%\sum_{\nu'_1 \nu'_2 l'} T^{(n)} _{\nu',\nu'_1\nu'_2l'}\,\left[\underbrace{ \varphi^{(n-6)}_{000}(A-1) \,\varphi_{0\nu'_2 l' }^{(n-3)}(\theta''_i,\hat{\xi}''_i)}_{\rm transformation \,(2)} \times\varphi_{K' \nu'_1 l' }^{(n)}(\theta'_1,\hat{\xi}'_1)\right]_{00}
%\eol
%=
%Y_0(1,...,A-2)\sum_{\nu'_1 \nu'_2 l'} \left( \delta_{\nu'_20}\delta_{l'0}+T^{(n)} _{\nu',\nu'_1\nu'_2l'}\right) \left[\varphi_{0\nu'_2 l' }^{(n-3)}(\theta_2,\hat{\xi}_2)\varphi_{K' \nu'_1 l' }^{(n)}(\theta'_1,\hat{\xi}'_1)\right]_{00}
%\eol
%+
%\sum_{i=2}^{A-1}Y_0(...,A-i-1,...) \,
%\sum_{\nu'_1 \nu'_2 l'} T^{(n)} _{\nu',\nu'_1\nu'_2l'}\,
%\sum_{\nu'_1 \nu'_2 l'} T^{(n-3)} _{\nu'_2 l',\nu''_1l''_1\nu''_2l''_2}\,
%\eol
%\times
%\left[ \left[\varphi^{(n-6)}_{0\nu''_2 l''_2}(\theta_i,\hat{\xi}_i) \times \varphi_{K''\nu''_1 l''_1 }^{(n-3)}(\theta_2,\hat{\xi}_2)\right]_{l'} \times\varphi_{K' \nu'_1 l' }^{(n)}(\theta'_1,\hat{\xi}'_1)\right]_{00}
\eeqn{ket}
The first term of this representation conveniently depends on the same vector $\ve{\xi}_2$ as the bra-state does. The second term of this expression can be easily re-expressed in the coordinates $\ve{\xi}_2$ and $\ve{\xi}'_1$ using  the same transformation (\ref{Transform1}), but  the third one, which depends on coordinates $\ve{\xi}''_3$, $\ve{\xi}''_2$ and $\ve{\xi}''_1$, shown in Fig. \ref{fig:coord-transform}, needs two transformations, also shown in this figure. 
%are schematically shown in Fig. \ref{fig:coord-transform}. 
The first transformation applies Eq. (\ref{Transform1}) to the product $\varphi^{(n-3)}_{000}(A') \,\varphi^{(n)}_{\nu'}(A-i)$ to express it in coordinates $\ve{\xi}''_i$ and $\ve{\xi}'_1$. 
\begin{figure}[t]
\vspace{0.5 cm}
\centering
\includegraphics[scale=0.37]{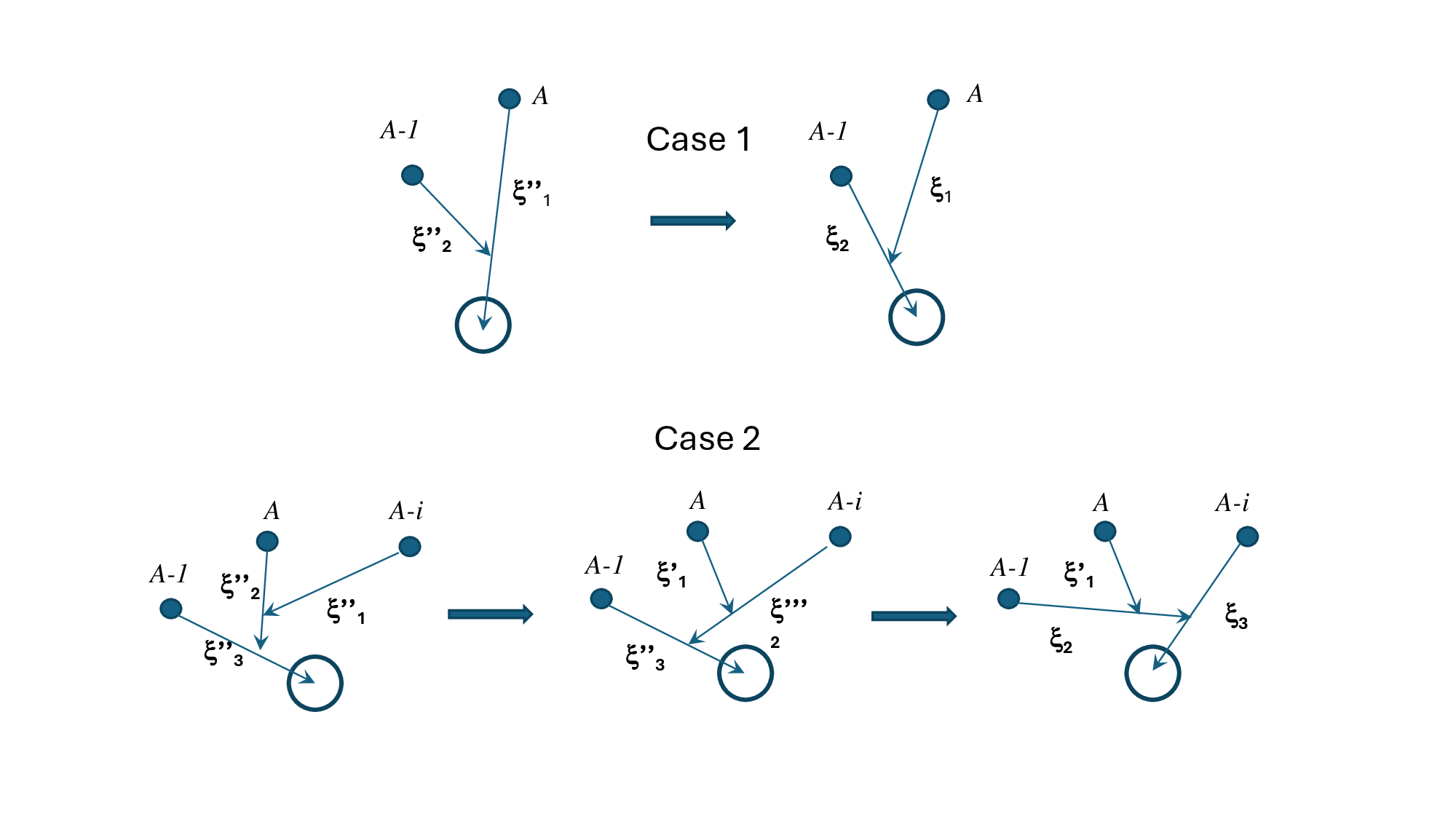}
\caption{Coordinates and their  transformations, employed in the bra- and ket-vectors of Eq. (\ref{MEcore}) shown as Case 1 and Case 2, respectively.}
\label{fig:coord-transform}
\end{figure}
Then   $\varphi^{(n-6)}_{000}(A-1) \,\varphi_{0\nu'_2 l' }^{(n-3)}(\theta''_i,\hat{\xi}''_i)$ is transformed as
\beq
\varphi^{(n-6)}_{000}(\theta''_3,\hat{\xi}''_3)  \,\varphi_{0\nu'_2 l'm' }^{(n-3)}(\theta''_i,\hat{\xi}''_i) 
=
\sum_{\nu''_1 \nu''_2 l''_1 l''_2} T^{(n-3)} _{\nu'_2 l',\nu''_1l''_1\nu''_2l''_2}\,
\eol
\times
 \left[\varphi^{(n-6)}_{0\nu''_2 l''_2}(\theta_i,\hat{\xi}_i) \times \varphi_{K''\nu''_1 l''_1 }^{(n-3)}(\theta_2,\hat{\xi}_2)\right]_{l'm'}, 
\eeqn{Transform2}
with $\gamma_1 = -1/(A-2)$ and $\gamma_2 = \sqrt{(A-1)(A-3)}/(A-2)$ and $2\nu''_2 + l''_2 +2\nu''_1+ l''_1 = 2\nu'_2+l'$. The coefficients $ T^{(n-3)} _{\nu'_2 l',\nu''_1l''_1\nu''_2l''_2}\,$ could be calculated from  recurrence relations derived in the next subsection. With these transformation the hyperangular matrix element  (\ref{OBME})  is deduced to
\beq
{\la 
\cal Y}_{\nu}(\hat{\ve{\rho}})
\mid
{\cal O}
\mid
{\cal Y}_{\nu'}(\hat{\ve{\rho}}')
\ra 
=
A^{-1}
\sum_{\nu_1\nu'_1lm} \frac{\delta_{\nu_2,\nu-\nu_1-l} }{2l+1} %\hat{l}^{-2} \,
{\cal C}^{(b)}_{\nu\nu_1\nu_2l}
{\cal C}^{(k)}_{\nu'\nu'_1\nu_2l}
\eol
\times\,
\varphi_{K' \nu_1 l m}^{(n)*}(\theta_1,\hat{\xi}_1) {\cal O}\, \varphi_{K \nu'_1 l m}^{(n)}(\theta'_1,\hat{\xi}'_1),
\,\,\,\,\,\,\,\,\,\,\,
\eeqn{OBME_HCM}
where
\beq
{\cal C}^{(b)}_{\nu\nu_1\nu_2l}
&=&
 {\cal N}_{\nu}^{-1/2}  
 \left(\delta_{\nu_2,0}\delta_{l,0} + (A-1)T^{(n)} _{\nu,\nu_1\nu_2l}\right)
 \eol
{\cal C}^{(k)}_{\nu'\nu'_1\nu_2l}
&=&
 {\cal N}_{\nu'}^{-1/2}  \left(\delta_{\nu'_20}\delta_{l'0}+T^{(n)} _{\nu',\nu'_1\nu_2l}  \, 
+
(A-2) T^{(n)} _{\nu',\nu'_1\nu'_2l}\,
 T^{(n-3)} _{\nu_2 l,\nu_2l00}
 \right). \,\,\,\,\,\,\,\,\,\,\,\,\,\,
\eeqn{Cb_Ck}

\subsection{Transformation coefficients for hyperangular functions}

In this section, the recurrence relations for transformation  coefficients $T^{(n)}_{\nu l, \nu l 0 0}$
\beq
T^{(n)}_{\nu l, \nu l 0 0} =
\la \varphi^{(n-3)}_{000 } (\theta'_2,\hat{{\xi}}'_{2})   \varphi^{(n)}_{0\nu l } 
(\theta'_1,\hat{\xi}'_{1})  \mid
\varphi^{(n-3)}_{000 } (\theta_2,\hat{{\xi}}_{2})   \varphi^{(n)}_{0\nu l } 
(\theta_1,\hat{\xi}_{1}) \ra, 
\eeqn{Tnul}
responsible for coordinate transformation (\ref{c-transform}) of hyperangular functions, are derived according to the scheme used in \cite{Tim23}.
 These coefficients are analogs of the Reynal-Revai coefficients for three-body systems \cite{RR}. 
%Similar to what is done in \cite{Tim23}, the recurrence relations for these coefficients are derived here. 
It is shown below that $T^{(n)}_{\nu+1 l, \nu+1 l 0 0}$ are related to 
\beq
 & T^{(n)}_{\nu l, \nu_1 l_1 n_2 l_2} (\gamma_1,\gamma_2) =
\eol
& \la 
[\varphi^{(n-3)}_{0n_2 l_2 } (\theta'_2,\hat{{\xi}}'_{2}) \times   \varphi^{(n)}_{2n_2+l_2 \, n_1 l_1} 
(\theta'_1,\hat{\xi}'_{1}) ]_{lm} \mid
\varphi^{(n-3)}_{000 } (\theta_2,\hat{{\xi}}_{2})   \varphi^{(n)}_{0\nu lm } 
(\theta_1,\hat{\xi}_{1}) \ra. \,\,\,\,\,\,\,\,\,\,\,\,\,\,\,
\eeqn{Tn}
The starting point of the derivation is  %the application to hyperangular function $\varphi^{(n)}_{0\nu lm } (\theta_1,\hat{\xi}_{1})$ 
the recurrence equation for hyperangular functions
\beq
\varphi^{(n)}_{K\nu+1 lm} 
(\theta,\hat{\xi}) &=&
\frac{{\cal N}^{(n)}_{K\nu+1 l}}{{\cal N}^{(n)}_{K\nu l}} a^{(n)}_{K \nu l} \varphi^{(n)}_{K \nu lm } (\theta,\hat{\xi})
+
\frac{{\cal N}^{(n)}_{K\nu+1 l}}{{\cal N}^{(n)}_{K\nu l}} b^{(n)}_{K\nu l} \cos 2\theta \varphi^{(n)}_{K\nu lm} (\theta,\hat{\xi})
\eol
&-&
\frac{{\cal N}^{(n)}_{k\nu+1 l}}{{\cal N}^{(n)}_{K\nu-1 l}} c^{(n)}_{K\nu l} \varphi^{(n)}_{K\nu-1 lm} (\theta,\hat{\xi}),
\eeqn{recrelphi}
where the coefficient $a,b,c$ are given by 
\beq
a^{(n)}_{K\nu l}& =&\frac{(2\nu+n/2+l-1)(n/2-l-3)(n/2+l-2)}{2(\nu+1)(\nu+n/2+l-1)(2\nu+n/2+l-2)}, 
\eol
b^{(n)}_{K\nu l} &=&
\frac{(2\nu + K+n/2 + l )(2\nu + K+n/2 + l-1) }{(\nu+1) (n+ 2\nu+2K   + 2l-2)},
\eol
c^{(n)}_{K\nu l} 
&=& \frac{ (2\nu+ n-5 )(\nu+l+1/2)(2\nu+n/2+l)}{2(\nu+1)(\nu+n/2+l-1)(2\nu+n/2+l-2)}.
\eeqn{bcoef}
 Only the second term from (\ref{recrelphi})  gives non-zero contribution to the transformation coefficient (\ref{Tn}). This term contains $\cos \theta_1$, which in new coordinates becomes
\beq
\cos 2\theta_1
%&=& 2\cos^2 \theta_1-1 = 2\,\,\frac{(\gamma_1 \vec{\xi}'_1 + \gamma_2 \vec{\xi}'_2)^2}{\rho^2}-1
%\eol
&=& 2 \gamma_1^2 \cos^2 \theta'_1 +4 \gamma_1 \gamma_2 \cos \theta'_1 \sin \theta'_1 \cos \theta'_2 \,(\hat{\vec{\xi}}'_1 \cdot \hat{\vec{\xi}}'_2)
\eol
& &+ 2 \gamma_2^2 \cos^2 \theta'_2 \sin^2 \theta'_1 - 1.
%\eol
\eeqn{cos2theta}
As in \cite{Tim23}, the transformation coefficient (\ref{Tn}) could be split into
\beq
T^{(n)}_{\nu+1 l ,\nu_1 l_1 \nu_2 l_2 }(\gamma_1,\gamma_2)
= (T.1) + (T.2) + (T.3),
\eeqn{Tsum}
according to the first three terms in (\ref{cos2theta}). The forth term  in (\ref{cos2theta}) gives zero contribution. 
Below, the following relations will also be needed:
\beq
(1\pm\cos 2\theta) \varphi^{(n)}_{K\nu lm} (\theta,\hat{\xi}) 
&=&
\pm A^{(n)}_{K\nu l} \varphi^{(n)}_{K\nu+1 lm} (\theta,\hat{\xi})+
(1\mp B^{(n)}_{K\nu l}) \varphi^{(n)}_{K \nu lm } (\theta,\hat{\xi})
\eol
&\pm&
C^{(n)}_{K\nu l} \varphi^{(n)}_{K\nu-1 lm} (\theta,\hat{\xi}),
\eeqn{1.pm.cos2t}
where
\beq
A^{(n)}_{K\nu l}= \frac{N^{(n)}_{K\nu l}}{b^{(n)}_{K\nu l}N^{(n)}_{K\nu+1 l}},
 \,\,\,\,\,\,\,\,\,\,\,
B^{(n)}_{K\nu l}= \frac{a^{(n)}_{K\nu l}}{b^{(n)}_{K\nu l}}, 
 \,\,\,\,\,\,\,\,\,\,\,
 C^{(n)}_{K\nu l} =\frac{ c^{(n)}_{K\nu l} }{ b^{(n)}_{K\nu l}} \frac{N^{(n)}_{K\nu  l}}{N^{(n)}_{K\nu-1 l}}. \,\,\,\,\,\,\,\,\,\,\,\,
 \eeqn{abc}
%\beq
%(1+\cos 2\theta) \varphi^{(n)}_{K\nu lm} (\theta,\hat{\xi}) 
%&=&
%A^{(n)}_{K\nu l} \varphi^{(n)}_{K\nu+1 lm} (\theta,\hat{\xi})+
%(1-B^{(n)}_{K\nu l}) \varphi^{(n)}_{K \nu lm } (\theta,\hat{\xi})
%\eol
%&+&
%C^{(n)}_{K\nu l} \varphi^{(n)}_{K\nu-1 lm} (\theta,\hat{\xi}),
%\eeqn{1+cos2t}
%\beq
%(1-\cos 2\theta) \varphi^{(n)}_{K\nu lm} (\theta,\hat{\xi}) 
%&=&
%-A^{(n)}_{K\nu l} \varphi^{(n)}_{K\nu+1 lm} (\theta,\hat{\xi})+
%\left[1+B^{(n)}_{K\nu l}\right] \varphi^{(n)}_{K \nu lm } (\theta,\hat{\xi}) 
%\eol
%&-&
%C^{(n)}_{K\nu l} \varphi^{(n)}_{K\nu-1 lm} (\theta,\hat{\xi}).
%\eeqn{one-sinus}

\subsubsection{Beginning of iterations}

The first term of the iterations is $T^{(n)}_{0l,\nu_1 l_1\nu_2 l_2}(\gamma_1,\gamma_2)$
\beq
T^{(n)}_{0l,\nu_1 l_1\nu_2 l_2}(\gamma_1,\gamma_2)
&=&
\int d\theta'_1 d\theta'_2 d\hat{\xi}'_1 d\hat{\xi}'_2 \,\,
[\varphi^{(n-3)}_{0\nu_2 l_2 } (\theta'_2,\hat{{\xi}}'_{2}) \times   \varphi^{(n)}_{2\nu_2+l_2 \, \nu_1 l_1} 
(\theta'_1,\hat{\xi}'_{1}) ]_{lm}
\eol
&\times&
\frac{1}{\sqrt{4\pi}} N^{(n-3)}_{000}N^{(n)}_{00l} \cos^l \theta \,\,Y_{lm}(\hat{\xi}_1).
\eeqn{T_0l}
Using
\beq
 \xi_1^l \,Y_{lm}(\hat{\xi}_1) = \sqrt{4\pi} \sum_{\lambda_1 +\lambda_2=l} 
 (-)^{\lambda_1} \sqrt{\frac{(2l+1)!}{(2\lambda_1+1)!(2\lambda_2+1)!}}
 %C_{\lambda_1 \lambda_2 l} 
 \eol
 \times
 \left[ (\gamma_2 \xi_2')^{\lambda_2} Y_{\lambda_2}(\hat{{\xi}}'_{2}) \times  (\gamma_1 \xi_1')^{\lambda_1} Y_{\lambda_1}(\hat{{\xi}}'_{1})\right]
 \eeqn{}
 %where
 %\beq
 %C_{\lambda_1\lambda_2 l} = (-)^{\lambda_1} \sqrt{\frac{(2l+1)!}{(2\lambda_1+1)!(2\lambda_2+1)!}}
%\eeqn{}
together with
 \beq
\cos^{\lambda_2}\theta'_2 \, Y_{\lambda_2 m_2}(\hat{{\xi}}_{2}) = \frac{\varphi^{(n-3)}_{00\lambda_2m_2}(\theta_2', \hat{\xi}'_2)}{N^{(n-3)}_{00\lambda_2}}
\eol
\cos^{\lambda_1}\theta'_1 \sin^{\lambda_1}\theta'_1\, Y_{\lambda_1 m_1}(\hat{{\xi}}_{1}) = \frac{\varphi^{(n )}_{00\lambda_1m_1}(\theta_1', \hat{\xi}'_1)}{N^{(n )}_{\lambda_2 0\lambda_1}}
\eeqn{}
it is easy to obtain
\beq
T^{(n)}_{0l,\nu_1 l_1\nu_2 l_2}(\gamma_1,\gamma_2)
%%%%%%%%%%%%%%%%%%%%%%
%%
%%   Uncomment for more detail
%%
%%%%%%%%%%%%%%%%%%%%%
%&=& 
%\sum_{\lambda_1 +\lambda_2=l} C_{\lambda_1 \lambda_2 l} \gamma_1^{\lambda_1}\gamma_2^{\lambda_2}\frac{N^{(n-3)}_{000}N^{(n)}_{00l}}{N^{(n-3)}_{00\lambda_2}N^{(n )}_{\lambda_2 0\lambda_1}}
%\eol
%&\times&
%\la[\varphi^{(n-3)}_{0\nu_2 l_2 } %(\theta'_2,\hat{{\xi}}'_{2}) 
%\times   \varphi^{(n)}_{2\nu_2+l_2 \, \nu_1 l_1} 
%%(\theta'_1,\hat{\xi}'_{1})
%]_{lm}\mid [\varphi^{(n-3)}_{00  \lambda_2 } %%(\theta'_2,\hat{{\xi}}'_{2}) 
%\times   \varphi^{(n)}_{\lambda_2 \, 0 \lambda_1} 
%%(\theta'_1,\hat{\xi}'_{1})
%]_{lm}\ra 
%\eol
%&=&\delta_{\nu_2 0}\delta_{\nu_1 0}\delta_{l_1+l_2,l}
%  C_{l_1 l_2 l} \gamma_1^{l_1}\gamma_2^{l_2}\frac{N^{(n-3)}_{000}N^{(n)}_{00l}}{N^{(n-3)}_{00l_2}N^{(n )}_{l_2 0l_1}}
%%  \eol  &=&\delta_{\nu_2 0}\delta_{\nu_1 0}\delta_{l_1+l_2,l} \gamma_1^{l_1}\gamma_2^{l_2} (-)^{l_1}  \sqrt{\frac{(2l+1)! \Gamma\left(l_1+\frac{3}{2}\right)\Gamma\left(l_2+\frac{3}{2}\right)}{(2l_1+1)!(2l_2+1)!\Gamma\left(\frac{3}{2}\right)\Gamma\left(l+\frac{3}{2}\right)}}
%%\eol
%%&=&\delta_{\nu_2 0}\delta_{\nu_1 0}\delta_{l_1+l_2,l}   \gamma_1^{l_1}\gamma_2^{l_2} (-)^{l_1}   \sqrt{2^{2l+1-2l_1-1-2l_2-1+1} \frac{l!}{l_1!l_2!} }
%\eol
&=&
(-)^{l_1}2^{l-l_1-l_2} \gamma_1^{l_1}\gamma_2^{l_2}
 \sqrt{ \frac{l!}{l_1!\,l_2!} }\, \delta_{\nu_2 0}\,\delta_{\nu_1 0}\,\delta_{l_1+l_2,l}. \,\,\,\,\,\,\,
\eeqn{iter1}

\subsubsection{Contribution from (T.1)}
With the help of the transformation (\ref{Transform2}) 
%\beq
%\varphi^{(n-3)}_{000 } (\theta_2,\hat{{\xi}}_{2})   \varphi^{(n)}_{0\nu l m} 
%(\theta_1,\hat{\xi}_{1}) 
%&=& \sum_{2\nu_1+2\nu_2+l_1+l_2=2\nu+l} T^{(n)}_{\nu l,\nu_1 l_1\nu_2 l_2} (\gamma_1,\gamma_2)
%\eol &\times& 
%[\varphi^{(n-3)}_{0\nu_2l_2 } (\theta'_2,\hat{{\xi}}'_{2})  \times \varphi^{(n)}_{2\nu_2+l_2 \, \nu_1 l_1 } 
%(\theta'_1,\hat{\xi}'_{1})]_{lm} .\,\,\,\,
%\eeqn{TC}
this contribution becomes
\beq
(T.1) &=&
%\gamma_1^2 \frac{{ N}^{(n)}_{0\nu+1 l}}{{ N}^{(n)}_{0\nu l}} b^{(n)}_{0\nu l}
%\sum_{2\nu'_1+2\nu'_2+l'_1+l'_2=2\nu+l} T^{(n)}_{\nu l,\nu'_1 l'_1\nu'_2 l'_2} (\gamma_1,\gamma_2)
% \,\,\,\,\,\,\,\,\,\,\,\,
%\eol
%\times
%\la [\varphi^{(n-3)}_{0\nu_2 l_2 } (\theta'_2,\hat{{\xi}}'_{2}) \times   \varphi^{(n)}_{2\nu_2+l_2 \, \nu_1 l_1} 
%(\theta'_1,\hat{\xi}'_{1}) ]_{lm} \mid   1+\cos 2 \theta'_1 
%\eol
%\times
%\mid
%[\varphi^{(n-3)}_{0\nu'_2l'_2 } (\theta'_2,\hat{{\xi}}'_{2})  \times \varphi^{(n)}_{2\nu'_2+l'_2 \, \nu'_1 l'_1 } 
%(\theta'_1,\hat{\xi}'_{1})]_{lm}\ra.
%\eol
 %=
\gamma_1^2 \frac{{ N}^{(n)}_{0\nu+1 l}}{{ N}^{(n)}_{0\nu l}} b^{(n)}_{0\nu l}
\sum_{2\nu'_1+2\nu'_2+l'_1+l'_2=2\nu+l} T^{(n)}_{\nu l,\nu'_1 l'_1\nu'_2 l'_2} (\gamma_1,\gamma_2) \delta_{\nu_2 \nu'_2} \delta_{l_2 l'_2}\delta_{l'_1 l_1}
 \,\,\,\,\,\,\,\,\,\,\,\,
\eol
& &\times
\la   \phi^{(n)}_{2\nu_2+l_2 \, \nu_1 l_1} 
(\theta'_1)  \mid   1+\cos 2 \theta'_1  \mid
 \phi^{(n)}_{2\nu'_2+l'_2 \, \nu'_1 l'_1 } 
(\theta'_1)\ra.
%\eol
\eeqn{Tnul}
Then  using (A.8) and (A.9) from \cite{Tim23} the (T.1) contribution   is  rearranged to  
\beq
(T.1) =
\gamma_1^2 \,\frac{{ N}^{(n)}_{0\nu+1 l}}{{ N}^{(n)}_{0\nu l}} 
\frac{N^{(n)}_{K'\nu_1-1 l_1}}{N^{(n)}_{K'\nu_1 l_1}}
\frac{b^{(n)}_{0\nu l} }{b^{(n)}_{K'\nu_1-1 l_1}}\,T^{(n)}_{\nu l,\nu_1-1 l_1\nu_2 l_2} (\gamma_1,\gamma_2). %\delta_{\nu'_1+1,\nu_1}
\eeqn{}

\subsubsection{Contribution from (T.2)}
Using
\beq
(\hat{\vec{\xi}}'_1 \cdot \hat{\vec{\xi}}'_2)[Y_{l_2 } (\hat{{\xi}}'_{2})  \times Y_{l_1 } (\hat{\xi}'_{1})]_{lm}=\hat{l}_1 \hat{l}_2 \sum_{\lambda_1 \lambda_2} (-)^{\lambda_2 + l_1 + l}\left\{
\begin{array}{lll}  l_2 & l_1 &l \\ \lambda_1 & \lambda_2 & 1\end{array}
\right\}
\eol
\times 
(1 0 l_1 0| \lambda_1 0) (1 0 l_2 0| \lambda_2 0) [Y_{\lambda_2 } (\hat{{\xi}}'_{2})  \times Y_{\lambda_1 } (\hat{\xi}'_{1})]_{lm}
\eeqn{}
the contribution from (T.2) becomes
\beq
(T.2)
&=&
%\frac{{ N}^{(n)}_{0\nu+1 l}}{{ N}^{(n)}_{0\nu l}} b^{(n)}_{0\nu l}
%\la [\varphi^{(n-3)}_{0n_2 l_2 } (\theta'_2,\hat{{\xi}}'_{2}) \times   \varphi^{(n)}_{2n_2+l_2 \, n_1 l_1} (\theta'_1,\hat{\xi}'_{1}) ]_{lm}\mid 4 \gamma_1 \gamma_2 \cos \theta'_1 \sin \theta'_1 \cos \theta'_2 
%\eol
%&\times&  (\hat{\vec{\xi}}'_1 \cdot \hat{\vec{\xi}}'_2)\sum_{2\nu'_1+2\nu'_2+l'_1+l'_2=2\nu+l} T^{(n)}_{\nu l,\nu'_1 l'_1\nu'_2 l'_2} %(\gamma_1,\gamma_2)[\varphi^{(n-3)}_{0\nu'_2l'_2 } (\theta'_2,\hat{{\xi}}'_{1}) \times\varphi^{(n)}_{2\nu'_2+l'_2 \, \nu'_1 l'_1 } (\theta'_1,\hat{{\xi}}'_{2})]_{lm} \ra
%\eol
%&=&
4 \gamma_1 \gamma_2\frac{{ N}^{(n)}_{0\nu+1 l}}{{ N}^{(n)}_{0\nu l}} b^{(n)}_{0\nu l}
\sum_{2\nu'_1+2\nu'_2+l'_1+l'_2=2\nu+l} T^{(n)}_{\nu l,\nu'_1 l'_1\nu'_2 l'_2} %(\gamma_1,\gamma_2)
\eol
&\times&
\, \la \varphi^{(n-3)}_{0n_2 l_2 } (\theta'_2)    \varphi^{(n)}_{2n_2+l_2 \, n_1 l_1} 
(\theta'_1) 
\mid
 \cos \theta'_1 \sin \theta'_1 \cos \theta'_2 \mid \varphi^{(n-3)}_{0\nu'_2l'_2 } (\theta'_2) 
\varphi^{(n)}_{2\nu'_2+l'_2 \, \nu'_1 l'_1 } (\theta'_1)\ra
\eol &\times& 
\hat{l}_1 \hat{l}_2   (-)^{l'_2 + l_1 + l}\left\{
\begin{array}{lll}  l_2 & l_1 &l \\ l'_1 & l'_2 & 1\end{array}
\right\}
(1 0 l_1 0| l'_1 0) (1 0 l_2 0| l'_2 0). 
\eeqn{T2}
Then   integration over $d\theta'_2$ is done using eqs. (A.15) and (A.21) from \cite{Tim23}, leading to
\beq
& \la & \varphi^{(n-3)}_{0 \nu_2 l}(\theta_2) \mid \cos \theta_2 \mid \phi^{(n-3)}_{0 \nu'_2 l'_2} (\theta_2) \ra
=
\frac{ N^{(n-3)}_{0\nu'_2 l'_2}}{N^{(n-3)}_{0\nu_2 l_2 }}\left(2\nu'_2+\frac{n-5}{2}+l'_2\right)^{-1}
\eol
&\times& \left[
\left(\nu'_2+l'_2+\frac{1}{2}\right)\delta_{l'_2-1, l_2} \delta_{\nu'_2 , \nu_2}  
+
\left(\nu'_2+\frac{n-5}{2}+l'_2\right)\delta_{l'_2+1, l_2} \delta_{\nu'_2 , \nu_2} 
\right.
\eol
& & + \left.
(\nu'_2+1) \delta_{l'_2-1, l} \delta_{\nu'_2+1, \nu_2}
+
\left(\nu'_2+\frac{n-8}{2}\right) \delta_{l'_2+1, l_2} \delta_{\nu'_2-1, \nu_2}\right].
\eeqn{ME_theta_2}
It is followed  by  integration over $d\theta_1$, resulting in
\beq
\la \varphi^{(n)}_{K'\nu_1 l} (\theta_1)\mid \cos \theta_1  \sin \theta_1 \mid \phi^{(n)}_{K'' \nu'_1 l'_1}  (\theta_1)\ra 
=\frac{N^{(n)}_{K'' \nu'_1 l'_1}}{N^{(n)}_{K'  \nu_1  l_1 }}
\,\,\,\,\,\,\,\,\,\,\,\,\,\,\,\,\,\,\,\,\,\,\,\,\,\,\,\,\,
\eol
\times \left[
\delta_{K'',K-1} \delta_{\nu'_1+1,\nu_1} \delta_{l'_1,l_1+1}
\frac{2\nu_1(n+4\nu +  2l  -2\nu_1)}{({n+4\nu + 2l-2 })(n+4\nu+2l )}  \right.
\eol
- \,\delta_{K'',K'+1}\delta_{\nu'_1+2,\nu_1}\delta_{l'_1,l_1+1}
%\frac{N^{(n)}_{K'' \,\nu'_1 \,l'_1}}{N^{(n)}_{K'  \nu_1 l_1 }}
\frac{4(\nu_1-1)\nu_1}{(  n +4\nu +2l )({ n+4\nu+2l-2})} 
\eol
+ \delta_{K'',K-1}\delta_{\nu'_1,\nu_1}\delta_{l'_1,l_1-1}
%\frac{N^{(n)}_{K'' \nu'_1 l'_1}}{N^{(n)}_{K'  \nu'_1  l'_1 }}
\left( 1 -\frac{2\nu_1}{n+4\nu+2l-2}\right)\left(1-\frac{2\nu_1}{n+4\nu +2l}\right)
\eol
-\left. \delta_{K'',K'+1}\delta_{\nu'_1+1,\nu_1}\delta_{l'_1,l_1-1}
%\frac{N^{(n)}_{K'' \nu'_1 l'_1}}{N^{(n)}_{K'  \nu_1  l }}
\frac{2\nu_1 }{n+4\nu+2l  -2}
\frac{n+4\nu+2l-2\nu_1 }{n+4\nu+2l }\right]
+ ..., \,\,\,\,\,\,\,\,\,\,\,\,\,\,\,
\eeqn{ME_theta_1}
where by $...$ the terms are denoted that do not satisfy $2\nu'_1+l'_1+2\nu'_2+l'_2=2\nu+l$ and    do not contribute to $(T.2)$.

\subsubsection{Contribution from (T.3)}

In a similar fashion one obtains the contribution from (T.3):
\beq
(T.3)
&=&
\gamma_2^2 \frac{{ N}^{(n)}_{0\nu+1 l}}{{ N}^{(n)}_{0\nu l}} b^{(n)}_{0\nu l}
\sum_{\nu'_1,\nu'_2} T^{(n)}_{\nu l,\nu'_1 l_1\nu'_2 l_2} (\gamma_1,\gamma_2) %\delta_{l'_1l_1}\delta_{l'_2l_2}
% \,\,\,\,\,\,\,\,\,\,\,\,
\eol
&&\times
\la \phi^{(n-3)}_{0\nu_2 l_2 } (\theta'_2) \mid   1+\cos 2 \theta'_2   
\mid \phi^{(n-3)}_{0\nu'_2l_2 } (\theta'_2)  \ra
\eol
& &\times
\la
\phi^{(n)}_{2\nu_2+l_2 \, \nu_1 l_1} 
(\theta'_1)  \mid  \sin^2 \theta'_1
\mid  \phi^{(n)}_{2\nu'_2+l_2 \, \nu'_1 l_1 } 
(\theta'_1)\ra.
\eeqn{T3}
With (\ref{1.pm.cos2t}) the integration over $d\theta'_2$ results in
\beq
\la \phi^{(n-3)}_{0\nu_2 l_2 } (\theta'_2) \mid   1+\cos 2 \theta'_2   
\mid \phi^{(n-3)}_{0\nu'_2l_2 } (\theta'_2)  \ra
=
A^{(n-3)}_{0\nu'_2 l_2} \delta_{\nu'_2+1, \nu_2}
\eol
+
(1-B^{(n-3)}_{0 \nu'_2 l_2})\delta_{\nu'_2, \nu_2}
+
C^{(n-3)}_{0 \nu'_2 l_2}\delta_{\nu'_2-1, \nu_2}.
\eeqn{}
The matrix element that involves integration over $d\theta'_1$ contains several terms but   only the following ones contribute to (T.4):
\beq
\nu'_2 +1 = \nu_2: \,\,\,\, &
\la \phi^{(n)}_{K'\, \nu_1 l_1} 
  \mid \sin^2 \theta'_1
\mid  \phi^{(n)}_{K'-2 \, \nu'_1 l_1 } \ra
=
%\eol
& \frac{(\nu_1 + K'+\alpha+l_1-\frac{1}{2})_2}{(2\nu_1 + K'+\alpha+l_1-\frac{1}{2})_2} 
\,
\eol
&  & \times\frac{N^{(n)}_{K'-2 \nu'_1 l_1}}{N^{(n)}_{K'  \nu_1 l_1}}\delta_{\nu'_1,\nu_1 }+...
\eol
\nu'_2 = \nu_2: \,\,\,\,\, &
\la \phi^{(n)}_{K'\, \nu_1 l_1}  \mid \sin^2 \theta'_1
\mid  \phi^{(n)}_{K' \, \nu'_1 l_1 } \ra
= &
-\frac{1}{2}A^{(n)}_{K'\nu'_1 l_1} \delta_{\nu'_1,\nu_1-1}+...\,\,\,\,\,\,
\eol
\nu'_2 -1 = \nu_2: \,\,\,\,\, &
\la \phi^{(n)}_{K'\, \nu_1 l_1}   \mid \sin^2 \theta'_1
\mid  \phi^{(n)}_{K'+2 \, \nu'_1 l_1 } \ra
= &
...+\frac{N^{(n)}_{K'+2  \nu'_1  l_1}} {N^{(n)}_{K' \nu_1 l_1}}\delta_{\nu'_1,\nu_1-2}\, 
\eol
& &\times 
\frac{4\nu_1(\nu_1-1)}{(n+4\nu+ 2l )(n+4\nu+2l -2)}. 
\eol
\eeqn{}

\begin{acknowledgements}
This work was supported by the United Kingdom Science and Technology Facilities Council (STFC) under Grant  No.   ST/V001108/1.
\end{acknowledgements}
%\bibliographystyle{aps-nameyear}      % American Physical Society (APS) style, author-year citations
%\bibliography{example}                % name your BibTeX data base
\nocite{*}


\end{document}